\documentclass[]{interact}
\usepackage{diagbox}
\usepackage{epstopdf}
\usepackage{natbib}

\usepackage[caption=false,subrefformat=parens,labelformat=parens]{subfig}
\usepackage{graphicx}

\usepackage{algorithm}
\usepackage{algorithmic}

\usepackage{natbib}
\bibpunct[, ]{(}{)}{;}{a}{}{,}
\renewcommand\bibfont{\fontsize{10}{12}\selectfont}

\theoremstyle{plain}

\theoremstyle{definition}

\theoremstyle{remark}

\usepackage[table]{xcolor}
\definecolor{lightsalmonpink}{rgb}{1.0, 0.6, 0.6}
\definecolor{atomictangerine}{rgb}{1.0, 0.6, 0.4}
\definecolor{pink-orange}{rgb}{1.0, 0.6, 0.4}
\definecolor{persianorange}{rgb}{0.85, 0.56, 0.35}
\definecolor{grannysmithapple}{rgb}{0.66, 0.89, 0.63}
\definecolor{darkseagreen}{rgb}{0.56, 0.74, 0.56}
\definecolor{lightgreen}{rgb}{0.6, 1, 0.6}
\usepackage{bbding}

\usepackage{amstext} 
\usepackage{array}   
\newcolumntype{L}{>{$}l<{$}} 

\begin{document}


\title{A Multi-Modal Network Equilibrium Model with Interacting Mobility Service Providers' Strategies}

\author{
\name{Claudia Bandiera\textsuperscript{a}\thanks{CONTACT C. Bandiera. Email: claudia.bandiera@uni.lu}, Richard~D. Connors\textsuperscript{a}, Francesco Viti\textsuperscript{a}}
\affil{\textsuperscript{a}Faculty of Science, Technology and Communication, 
University of Luxembourg, Luxembourg}
}

\maketitle

\begin{abstract}
A Mathematical Program with Equilibrium Constraints (MPEC) is formulated to capture the relationships between multiple Mobility Service Providers (MSPs) and the users of a multi-modal transport network. The network supply structure is defined through a novel supernetwork approach where users’ daily trip chains are represented to model the mobility services used to reach each destination. At the upper level, a profit maximization formulation is introduced to describe each MSPs’ behaviour. At the lower level, users within a class choose minimum cost routes, according to Wardrop’s first equilibrium principle. To consider the interactions between modes, non-separable costs between supernetwork links are defined, and users’ equilibrium conditions are formulated as a Variational Inequality (VI). To solve the MPEC, an iterative solution algorithm based on a Modified Projection Method is proposed. Numerical examples are presented to illustrate properties of the model, and to examine scenarios showcasing cooperation or competition strategies between MSPs.
\end{abstract}


\section{Introduction}
Transportation systems have been offering an increasing number of multi-modal options thanks to the introduction of new mobility solutions (e.g., car- and bike-sharing, carpooling, e-scooter, on-demand services), often integrated into single platforms (Mobility-as-a-Service) with the purpose of improving the efficiency of last-mile connectivity to public transport and in turn to reduce private car use. The proliferation of these new services has increased the adoption of competition and cooperation strategies in the transportation market, where different mobility service providers (MSPs) operate in pursuit of both profits and business sustainability. Each MSP tries to  generate revenues by applying different policies, for instance by changing prices, varying their fleet size, or cooperating with other mobility suppliers to attract a sufficient share of the market demand. In this context, users’ modal choices play a fundamental role in determining the durability of mobility services inside the transportation system. Therefore, it is essential to design an appropriate transportation modelling approach to predict how users will make choices and react to MSPs’ strategies. Studying such interactions helps to understand in which conditions an MSP can have a profitable business, what happens when a competitor enters the market, or if cooperation can improve their value in the network.  

The problem of studying the interactions between transport operators and travellers has been widely studied in the literature in the context of uni-modal networks \citep{Farahani2013ARO}. When this interaction is analysed in a network where several modes of transport coexist, the problem can be defined as the multi-modal Network Design Problem (MNDP) \citep{Montella2000MultimodalND}. MNDPs are well known for their modelling complexity, arising from the structure of the multi-modal network, the definition of travel demand and users’ modal choices, and the formulation of the interaction between suppliers and demand. 

According to \cite{Nes2002DesignOM} there are three options to reduce the complexity of the representation of a multi-modal network. (i) Using a combination of a reduced number of modes of transport \citep{Liu2015SimultaneousOO}; (ii) focusing on a simple case, e.g. a transport corridor \citep{Tirachini2014MultimodalPA}; or (iii) using a hypernetwork \citep{Sheffi1978HYPERNETWORKSAS} (or supernetwork \citep{Sheffi1985UrbanTN}) approach, where the multi-modal network is divided in several uni-modal layers connected by transfer links \citep{Carlier2003ASA}. For the purpose of this study, attempting to represent a realistic and exhaustive multi-modal network, a supernetwork approach is taken into account to consider future scalability of the problem. 

Modelling the travel demand considering the activities that users perform at different destinations is fundamental in multi-modal network design problems \citep{LIU201524}. Hence to predict transport demand, it is of paramount importance to model trip chains and their impact on mode choices on a daily horizon. Supernetworks have been extended to include users’ daily trip chains in multi-modal contexts within activity-based models \citep{Arentze2004MultistateSA,Liao2010SupernetworkAF} and in multimodal network assignment \citep{Fu2014ANE,Liu2020DaytodayNA}. These models are based on the assumption that, during a day, each travel choice made by users is influenced both by earlier decisions and by planned later trips \citep{scheffer2021trip}. Following these assumptions, this study seeks to develop a general formulation of the MNDP, using a representation based on a supernetwork approach and that considers sequential mode choices determined by users' trip chains. Due to the complexity of the problem caused by the combinatorial explosion of trip chain options in time and space, and since we aim for developing a strategic long-term equilibrium model, our methodology is applied to a static system in which time of departure/arrive from/to a location or duration of the activities performed at each destination are not explicitly considered. The proposed approach is therefore intended for strategic economic planning and assessment applications, rather than for operational, management problems.

The complexity connected with the travel demand definition, however, still increases when users' heterogeneity is taken into account. Following \cite{hasan2007multiclass}, to represent a more realistic system users are divided into classes based on their socio-economic attributes and their trip chains. Reproducing the heterogeneity of users' choices is fundamental to properly determine path costs and to perform assignment procedures \citep{Halat2016DynamicNE}. Generally, assignment models are used to represent users’ behaviour in the transportation network. These models differ due to their assumptions regarding demand, supply and route choice \citep{book}, along with the assumed state of the system (e.g., equilibrium) and the representation of time (static or dynamic). In accordance with the strategic application purpose, in this paper, we adopt a static traffic assignment approach, which computes the steady-state equilibrium condition considering constant demand and supply. The canonical model of this type (review in \cite{Saw2015LiteratureRO}) is the user equilibrium condition, established by \cite{Wardrop1952SomeTA}, where at equilibrium all used paths have equal minimum cost so that no user has any incentive to change path. In their seminal paper, \cite{Beckman1956StudiesIT} formulated the user equilibrium link flows as the solution to a minimisation problem \citep{Boyce2005ARO}, making it solvable with various optimisation algorithms. Unfortunately, in the case of non-separable link cost functions one could no longer reformulate the network equilibrium conditions as a solution to an optimisation problem \citep{Boyce2005ARO}. To deal with the non-separable case, \cite{Smith1979TheEU} and \cite{Dafermos1980TrafficEA} expressed the user equilibrium in the form of a variational inequality (VI) and subsequently a nonlinear complementary problem (NCP) formulation was provided by \cite{Aashtiani1981EquilibriaOA}. In this paper, we define the lower-level equilibrium conditions with fixed demand in the form of VI considering non-separable link cost functions, based on the assumption that multiple user classes travelling with different modes of transport in the same network do influence each other.

Finally, in an MNDP the interaction between suppliers and users is often modeled as a bi-level problem, i.e. an optimization problem characterised by an optimization problem as a constraint at the lower level \citep{Sinha2017ARO}. In this two-level decision-making process, at the upper level a leader sets some directives that are going to influence the decisions of other agents, called followers, situated at the lower level \citep{Luo1996MathematicalPW}. More specifically, the problem architecture usually considers that an operator controls some network parameters/strategies seeking to maximise profits, sometimes in combination with other objectives (e.g., ensuring basic service level to all users). In response, users change their travel choices to maximise their utility, generally equivalent to minimising time and monetary costs. Generally, these types of models typically result in non-convex optimisation problems, for which globally optimal solutions are usually difficult to establish. 
When, in a bi-level problem, the lower level is defined through equilibrium conditions in the form of VI or complementarity conditions, the problem is called Mathematical Program with Equilibrium Constraints (MPEC). According to \cite{Luo1996MathematicalPW} the MPEC analytically reproduces a Stackelberg game \citep{stack}, that can be considered as an extension of the Nash game \citep{Nash1951NONCOOPERATIVEG}. In the latter all players are considered at the same level, unilaterally trying to minimize their costs, and in order to achieve this purpose each of them is going to make a choice considering the one of the others. In Stackelberg games, instead, there is a more powerful actor, the leader, which can choose their actions anticipating the followers’ reactions, and thanks to this information decides the best strategy to adopt \citep{Luo1996MathematicalPW}. This game-theoretical approach has been already proposed in supplier-demand equilibrium problems such as road pricing or traffic signal optimisation. In network design modelling, the leader objective function can be defined through different formulations depending on the considered objective, for example as maximization of the total social welfare \citep{Fan2014BilevelPM}, minimization of the total travel time in the network \citep{Ye2018OptimalDO}, maximization of the total profit of the supplier \citep{Kalashnikov2016AHA}, or for jointly optimizing multi-modal transfers by defining location, capacity, and parking fees of park and ride facilities \citep{Ye2021JointOO}. In this study, to analyse the interaction between an MSP and transportation users, we formulate an upper-level profit maximisation in which the decision variable is the capacity of the service. Moreover, this formulation is built with a general structure that can be separately applied to different MSPs. 

The problem of optimally designing multi-modal networks has been tackled in various ways in recent years. In the context of multi-modal transit networks, \cite{Wan2009CongestedMT} use a State-Augmented Multi-modal (SAM) network (developed by \cite{Lo2002ModelingCM}) in which they could include various factors, such as road congestion for transit mode, users' multi-modal choices, in-vehicle crowdedness and  non-linear fare structures. In the proposed bi-level problem the authors try to maximize the social welfare of the system while optimizing network configuration and service frequency. Moreover, trough SAM they divided transport modes in classes to consider different levels of congestion. \cite{Cipriani2006AMT}, instead, developed a model in which they minimize costs, resources and externalities for a transit operator and users of a multi-modal transport system that includes private car. Buses routes and frequencies are the output of the model in which the transit travel time is influenced by transit characteristics and car travel times. However, travellers by private car are not experiencing a variation of travel time based on the presence of transit vehicles. A study conducted by \cite{Rashidi2016OptimalTC} includes a pedestrian transportation system as an independent mode of transport together with public transit and car. They developed a bi-level mathematical programming in order to determine the optimal location of sidewalks and crosswalks, while minimizing the total transportation cost and improving pedestrians’ safety. \cite{Fu2020MaximizingSA} developed an integration between Activity-Time-Space (ATS) network and the SAM transport network in the context of multi-modal transit networks, in which the individuals’ accessibility to different activities and travels are considered. The model is formulated as a bi-level programming problem, in order to maximize the activity-based space-time accessibility of activity locations. At the lower level individuals are considered as homogeneous, and their choices in terms of activities, departures, modes and routes are analysed. 

Although an MNDP considers the coexistence of different MSPs in the same multi-modal network, upper-level formulations are built to represent a specific MSP. However, the existing approaches are not general enough to be applied for diverse MSPs. Furthermore, in most cases, the modes of transport considered do not include all available services of the area, primarily focusing on transit or car systems. Moreover, these studies do not fully take into account the non-separability of the problem in which the congestion level of the different modes of transport are influenced by each other.

With the increased uptake of new shared mobility services, recent works have tried to introduce such options into network modelling problems. \cite{Nair2014EquilibriumND} defined a transit network in the presence of sharing systems. In this context, they defined a bi-level mixed-integer program, where at the upper level an MSP adjusts their decisional variables in order to maximize profit. At the lower level, instead, users minimize their travel times and waiting times. The authors converted the lower level into Karush–Kuhn–Tucker (KKT) conditions in order to turn the problem into an MPEC. In the context of bike-sharing systems, \cite{Caggiani2020AnEM} defined a model to identify the stations' location connected to public transport networks while considering equity. The authors minimize the accessibility to the transport system among a population divided in two groups based on socio-economic characteristics. A work of \cite{Nguyen2022AUA}, instead, developed a bi-level model considering a one-way car-sharing service in a multi-modal dynamic network using an activity-based approach. Travellers can be private car drivers, car-sharing drivers or transit passengers and perform three activities: go to work, shopping and return home. At the upper level the car-sharing operator tries to maximize their profit by controlling the price of the service and how to organize the vehicles at the depots. The lower-level activity-based choice model is written as a VI. Recently, \cite{NAJMI2023103959} introduced a multi-modal multi-provider market equilibrium model categorised as strategic modelling. Through this approach the authors assess the impact of different scenarios examining "what if" questions, without taking into account travellers' micro and meso-decisions. The problem is formulated as a Nash-Cournot model, in which a network operator, ride-sourcing providers and different classes of travellers are considered.  

The cited works try to capture the dynamics between users assigned in a multi-modal network and MSPs setting different strategies. However, to the authors' knowledge, there are no studies that formulate a general upper-level objective function that is sufficiently flexible to accommodate the diverse range of mobility services, while capturing the complex multi-modal trip chains of multi-class users, whose choices are influenced by congestion and limited service capacities. The focus of this paper is to develop a model able to study the economic assessment of any MSP influenced by the fixed strategies of other competitors and by users' choices in a multi-modal network. The decision variables of the model are the number of vehicles that the MSP can introduce in the network (their service capacity) and the path flows of users that represents the level of usage of each mobility service included in the supernetwork. Along with the listed characteristics of the model, the main contribution of this paper is based on the introduction in this type of problem of a lower level multi-class user equilibrium in which the different costs perceived by users are considered to be not-separable. This aspect increases drastically the complexity of the model, making the MPEC hard to be solved as an optimization problem using conventional solution algorithms. To solve the problem we propose an iterative solution algorithm: we optimize the upper level objective function through a sequential quadratic programming method, that solves a quadratic sub-problem at each iteration. Subsequently, for each upper level solution, we evaluate the path flow at equilibrium at the lower level trough the Modified Projection Method (MPM) \citep{Nagurney1992NetworkEA}.

The rest of the paper is structured as follows. After introducing the notation in Table \ref{table 1}, Section 2 describes the methodology, where the network structure and assumptions are explained, followed by the formulations of MSPs and user classes. Section 3, defines the general MPEC formulation and the proposed solution algorithm. Examples with numerical results are illustrated in Section 4. Finally, Section 5 discusses conclusions and potential developments of the methodology.

\begin{table}
\tbl{Notation}
{
\begin{tabular}{p{0.1\textwidth} p{0.4\textwidth}|p{0.15\textwidth} p{0.4\textwidth}} 
\toprule
 \textbf{Sets} & Description &\textbf{Indices} & Description\\ \midrule
$K$  & Set of user classes & $j$ & the MSP taken into account\\ 
$S$  & Set of mobility subscriptions  & $k$  & a user class\\ 
$A$  & Set of links  &$s$  & a mobility subscription\\
$A^{j}$  & Set of modal-links owned by the MSP & $a$  & a link of the network \\
$A_{s}$  & Set of subscription links& $w$  & an OD pair\\
$A_{s}^{j}$  & Set of subscription links involving the MSP& $p$  & a path connecting an OD \\
$N$  & Set of nodes  & &\\
$W$  & Set of OD pairs  & &\\
$P$  & Set of paths in the network  & & \\
$P_w$  & Set of paths between $w\in W$ &&\\
$\Phi$  & Set of all feasible path flows & & \\
$D$  & Total travel demand of the network & & \\ \midrule
 \textbf{Users' variables} & Description &\textbf{Supplier's variables} & Description\\ \midrule
$\bf{f}$  & vector of link flows & $v$ & fleet size of the MSP\\ 
$f_a$ & flow on link a & $v_a$ & number of vehicles on link $a\in A^{j}$ \\
$f^{k}_a$ & flow of class k on link a & \\ 
$\bf{x}$ & vector of path flows && \\
$x_p$ & flow on path p&&\\
$x^{k}_p$ & flow of class k on path p &&\\ \midrule
 \textbf{Parameters} & Description &\textbf{Functions} & Description\\ \midrule
 $d^{k}_w$ & demand of class $k$ on OD $w$ &$c_{lease}(v)$ & leasing cost (€)\\
 $\gamma$ & relocation factor & $t_{a,access}(f_a, v_a)$ & access time (hour)\\
 $\omega$ & booking factor& $t_{a,main}(\bf{f})$ & time in the main mode of transport (hour)\\
 $l_{a}$ & length of link $a$ (km)&$t_{a,egress}(f_a, v_a)$ & egress time (hour)\\
 $c_{s}$ & daily cost for subscription $s$ (€/day) &$t_{a,wait}(f_a, v_a)$ & waiting time (hour)\\
 $r_{s}$ & daily subsidy based subscription $s$ (€/day)& $t_{a,park}(f_a, v_a)$ & parking time (hour)\\
 $c_{a,h}$ & cost per hour $h$ travelling on link $a$ (€/hour)&$C_{a}^{k}(\textbf{f}, v_a)$ & total cost on link $a$ for class $k$\\
 $c_{a,km}$ & Cost per kilometre km on link a  (€/km) &$C_{a, access}^{k}(f_a, v_a)$ & total access cost on link $a$ for class $k$\\
$c_{a,fixed}$ & ticket cost on link $a$ (€/day) &$C_{a, main}^{k}(\textbf{f}, v_a)$ & total travel cost on link $a$ for class $k$\\
$c_{a,fuel}$ & fuel/recharge cost for vehicle (€/km) &$C_{a, egress}^{k}(f_a, v_a)$ & total egress cost on link $a$ for class $k$\\
$c_{a,park}$ & cost to find a parking slot  (€/hour) &$C_{p}^{k}(\bf{x}, v)$ & total travel cost on path $p$ for class $k$\\
$\delta_{a,p}$ & incidence matrix link-path &&\\
$\delta_{a,s}$ & incidence matrix link-subscription  && \\ \bottomrule
\end{tabular}}
\label{table 1}
\end{table}

\section{Methodology}

This section describes the methodological approach used in the paper. In Section 2.1 we define the assumptions connected to the structure of the network. Then, in Section 2.2 we describe the assumptions related to MSPs, user classes, and their formulations.

\subsection{Structure of the network and assumptions} \label{Sec2.1}
We introduce a multi-modal network approach in which users’ daily trip chains are explicitly modelled and where different mobility services available in the area are included. Since our interest is to study the long-term impacts, e.g. in terms of profitability of a certain service and how it can be optimised,
we keep the model at a regional/zonal scale, without detailing performance at the route level or for each individual. Hence, in our supernetwork approach we focus on aggregated users’ modal choices, and identify an MSP’s best fleet strategy in order to maximize the profit for a company serving the area of analysis.

\begin{figure}
\centering
\includegraphics[scale=0.1]{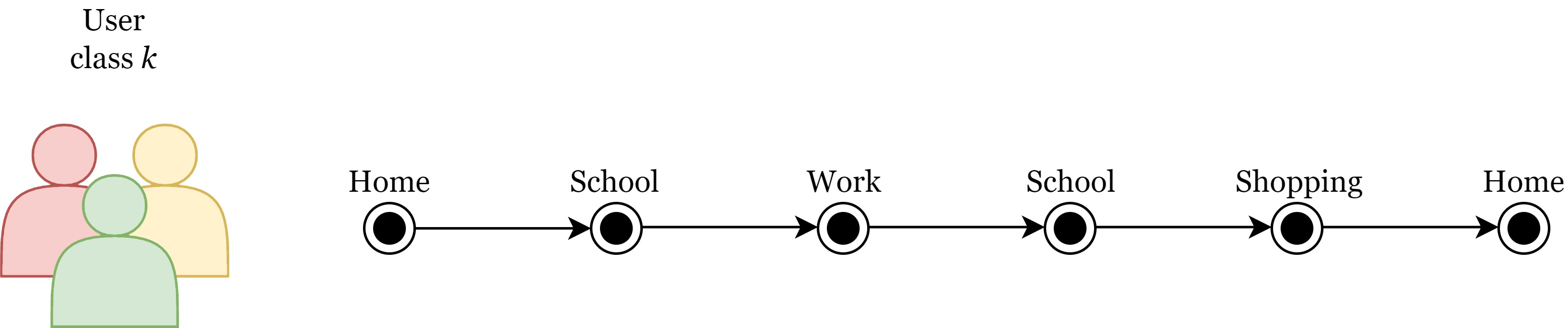}
\caption{User Class k with daily trip chain} \label{Fig1}
\end{figure}

\begin{figure}
\centering
\includegraphics[scale=0.1]{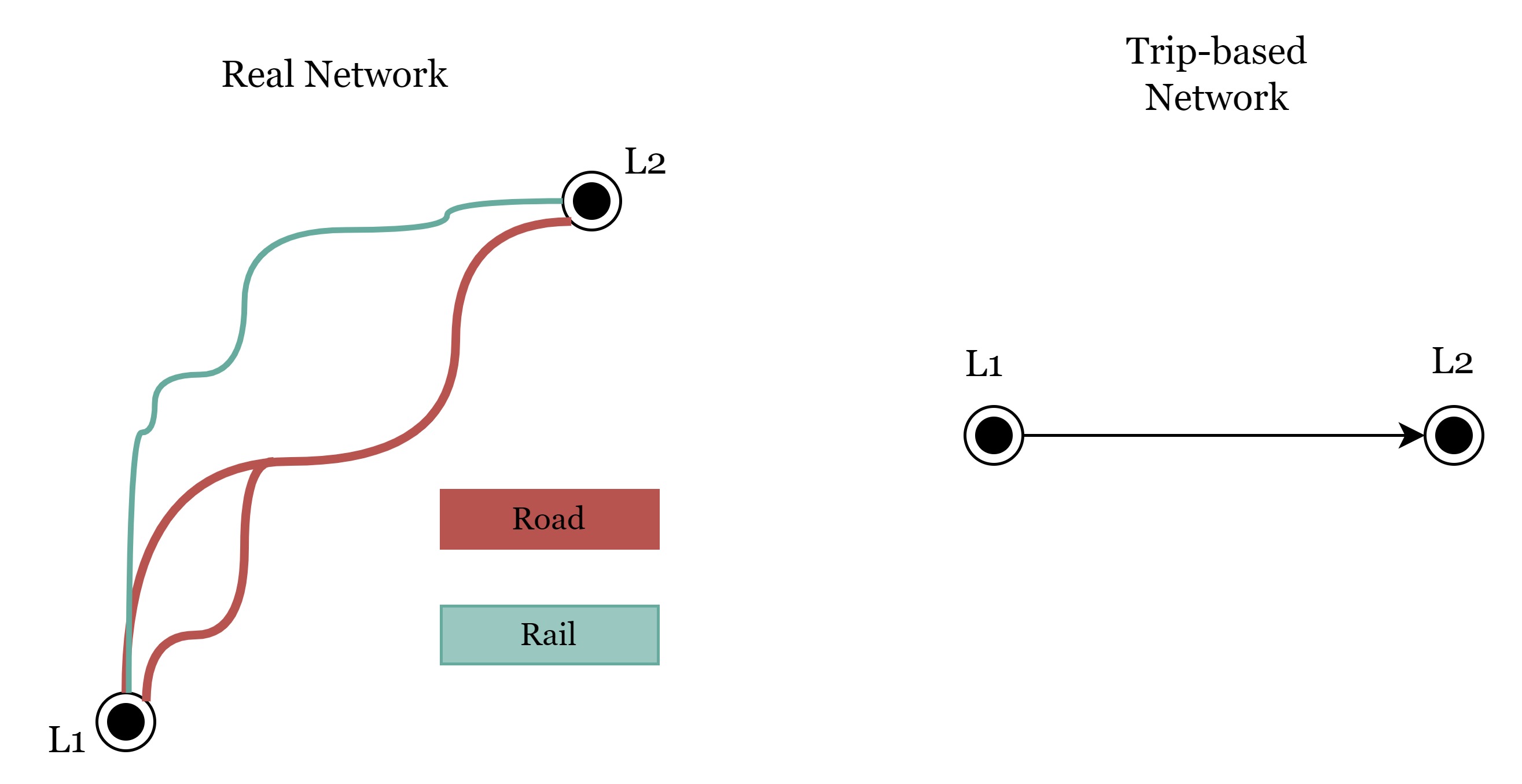}
\caption{Network assumption}
\label{Fig2}
\end{figure}

\begin{figure}
\centering
\includegraphics[scale=0.1]{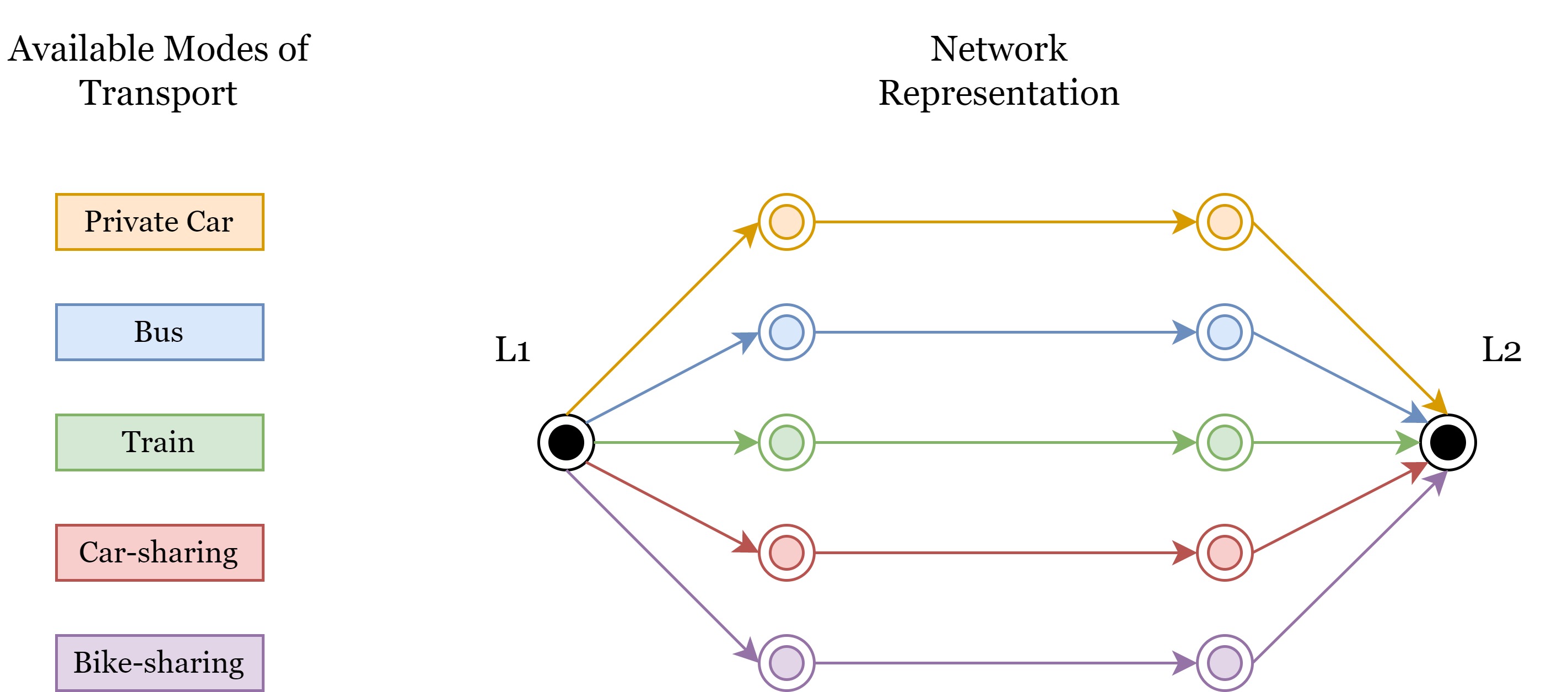}
\caption{Network expansion} 
\label{Fig3}
\end{figure}

\begin{figure}
\centering
\includegraphics[width=\textwidth]{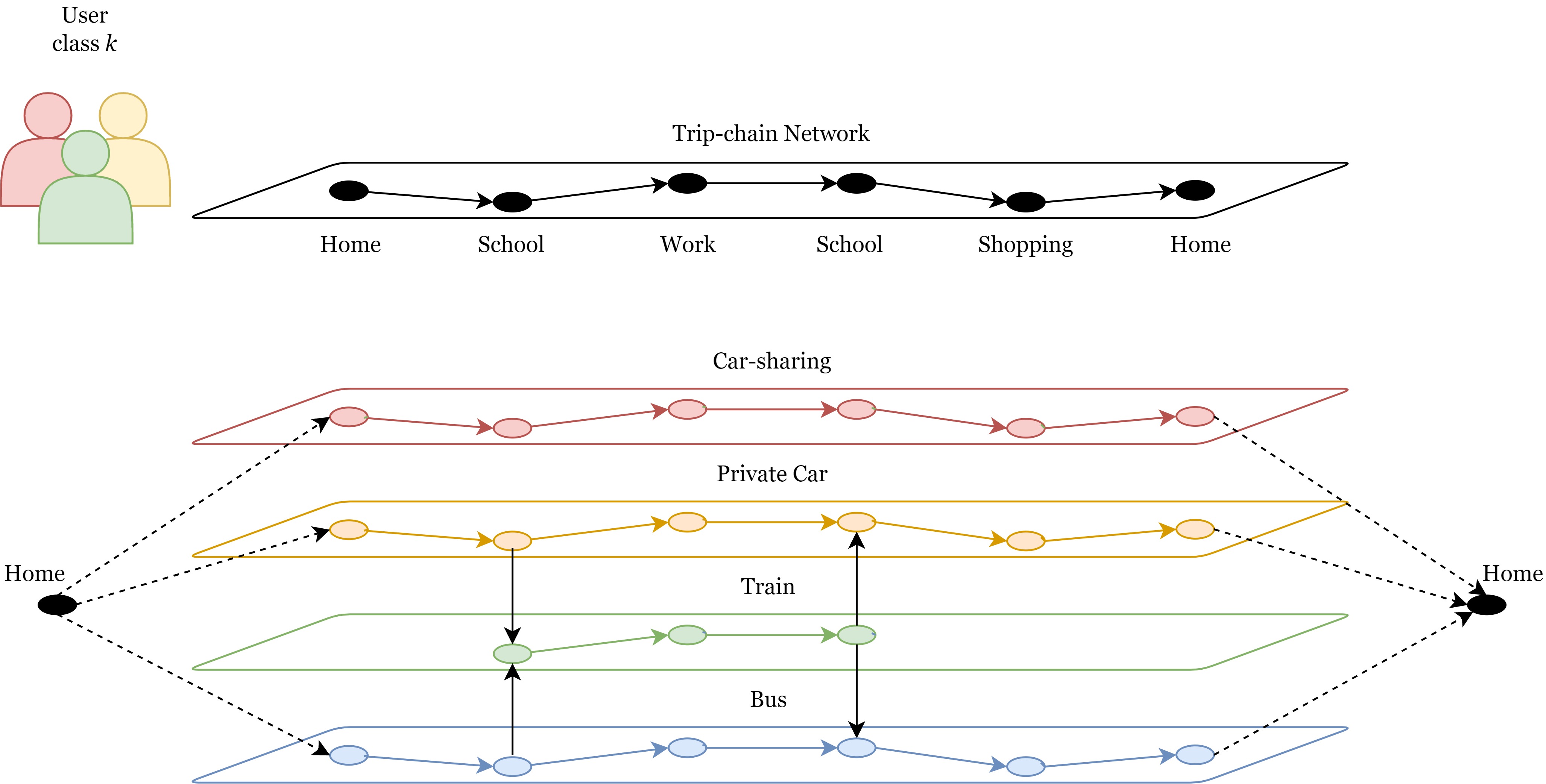}
\caption{Multi-modal Trip-chain Supernetwork} \label{Fig4}
\end{figure}

Let travellers be divided into $K$ classes based on their personal attributes and daily trip chains. We further assume that, during an ordinary weekday, users of the same class perform their activities in specific zones within the study area. The sequence of these trips is then modelled as a directed graph (Figure \ref{Fig1}) where a node $n$ corresponds to a location (zone) and a link $a$ indicates the trip from one location to another. In this example, users in class $k$ leave home to drop off their children at school, go to work, pick up their children for shopping, and finally return home together. For illustrative purposes, in Figure \ref{Fig1} the different visited locations of class $k$ are explicitly identified with the activities performed at each zone. However, in the model definition and application we do not require to explicitly model activities in detail; on the contrary, we consider that users of different classes can perform different activities in the same location, but a user class is simply determined by the specific sequence of visited locations, which clearly reduces the problem complexity. 
Moreover, this graph does not explicitly require to represent the real underlying transport infrastructure in detail. Specifically, a single link represents a trip without defining all the different travel alternatives available in the real network (Figure \ref{Fig2}). We assume that the network reaches equilibrium for which all used mode and route alternatives will reach the same generalised cost \citep{Wardrop1952SomeTA}, allowing us to represent them by a single aggregated link \citep{connors}. Subsequently, in order to explicitly represent the different modes of transport available between two locations and eventually operated by different MSPs, we expand the trip chain network in several uni-modal MSP-specific networks (Figure \ref{Fig3}). The first and last locations of the trip chain are also considered as centroids of the network, in order to guarantee flow conservation. As a result, with this representation we can take into account all the possible modes of transport active in an area, including users’ private vehicles, i.e. car, bike, scooter. Combining users’ trip chains with MSPs’ information the network in Figure \ref{Fig1} becomes a multi-modal trip chain-based supernetwork as shown in Figure \ref{Fig4}. 

According to the network representation illustrated in Figure \ref{Fig4}, each traveller must choose a path $p$ through the multi-modal network in order to perform the sequence of trips during their ordinary day. A path (representing a trip chain for the specific user class $k$) can comprise three different types of links: 1) access links (black dashed links) allow users to access a mode of transport from their origin (Home on the left side), and egress from a mode of transport to reach their final destination (Home on the right side). Another important role of the access links is that they can capture a subscription package needed for accessing a specific mode of transport (e.g., a public transport subscription) or the combination of integrated MSPs’ services (a Mobility-as-a-Service bundle subscription). 2) Mode-specific links (horizontal links) indicate trips made from one activity location to another using a specific mode of transport (designated by colour). 3) Interchange links (vertical black links) instead allow users to change to another transport mode when departing from an activity location. Each (within-layer) mode-specific link includes in turn three stages of a trip. In the first stage, the users leave the activity location and walk to reach the main mode of transport. Subsequently, travellers use the specific mode of transport in order to reach their next destination. Finally, the users leave that transport mode and walk to arrive at their next location \citep{Arentze2013TravelersPI}. Once they have performed all their trips, users return home completing their daily trip chain. 

We assume that each MSP seeks to maximize the profit generated from their service. This profit is calculated considering that a generic MSP $j$ manages specific and uniquely defined layers of the supernetwork, through which they collect revenues based on how many travellers use their service, while facing costs that depend on the service capacity provided e.g. number of vehicles. Hence, the studied MSP owns a fleet of vehicles $v$, and as part of their choice, at equilibrium, they will strategically distribute these vehicles across the links of their layer. The number of vehicles assigned at equilibrium to a specific link ($v_a$) will be considered available only for that trip connection. Due to the static assumption of the network, the daily (re-)location of vehicles is  represented by a coefficient in the fixed costs, assuming that the optimal fleet size depends only marginally on the actual position in time and space of the vehicles. The lower-level equilibrium decision variables, instead are the vector of path flows $\textbf{x}$.

\subsection{Model formulation and assumptions}
In this section, the previously introduced interaction between an MSP and users in a multi-modal trip-based supernetwork is formulated as an MPEC, considering an ordinary weekday. We first introduce a generic MSPs’ formulation, considering that their market choices are made based on a profit maximization objective. Subsequently, traffic network equilibrium conditions are proposed. 
\subsubsection{Mobility service providers and profit maximisation}
As a basic principle in economic theory, profit reaches its maximum value when the marginal revenue of a firm is equal to the marginal cost that must be sustained to offer the service \citep{Primeaux1994ProfitMT}. In order to represent the MSPs’ behaviour in the transport market, we considered that each of them seeks to maximize the profit arising from their own mobility service. MSP’s profit ($Pr$) is defined as the difference between total revenues (comprising fixed plus variable revenues) and total costs (including fixed plus variable costs):
\begin{equation}
Pr = TR - TC
\label{Eq1}
\end{equation} 

We now define the different revenue and cost components for a generic MSP, considering that the formulation can be applied for different modes of transport such as bike sharing, public transport, carpooling, car-sharing, e-scooter sharing, taxi and train. Depending on the type of service analysed some components could be omitted for the calculation, as shown in Table \ref{table 2}. Each unitary cost considered in this section depends on the capacity of the service. 

\paragraph{Revenues}We consider an MSP $j$ selling a mobility service to have two main revenue streams. The first component represents the fixed revenues ($FR$):
\begin{equation}
FR(\mathbf{f}) = \sum_{s\in A^{j}_s}\sum_{a\in A^{j}}(c_s + r_s)f_a
\label{eqfr}
\end{equation} 
where $a$ and $s$ indicate respectively a link of the network and a subscription to a mobility service or package.
In Equation \ref{eqfr}, based on the number of users subscribing to the service $f_a$, the MSP receives a fixed revenue from the subscription fee to access the service $c_s$ and from the potential subsidy $r_s$ coming from an external actor, i.e. the government or a local authority. 
The second component is represented by variable revenues ($VR$). This value depends on travellers’ usage of the service:
\begin{equation}
VR(\mathbf{f}) = \sum_{a\in A^{j}}(c_{a,h}t_{a,main}(\mathbf{f}) + c_{a,km}l_a + c_{a,fixed})f_a
\end{equation} 
The first term represents the time spent on board $t_{a,main}(\mathbf{f})$  using the main mode of transport. This component is non-separable, therefore depends on the link flows of the modes of transport available on the same trip connection. It is important to clarify that not all modes of transport necessarily contribute to congestion, e.g. trains have a separate infrastructure or buses can have dedicated lanes. This term is multiplied by the revenue related to time use $c_{a,h}$. The second term $l_a$ indicates distance travelled, multiplied by revenue per kilometre $c_{a,km}$. The third term $c_{a,fixed}$ is the revenue from a fixed fee/ticket imposed on the user each time the service is used.

\paragraph{Costs} An MSP experiences costs that can be also divided into fixed and variable costs. The fixed costs ($FC$) are defined in Equation \ref{Eq4} through a function that varies with the number of vehicles the supplier deploys on the network $c_{lease}(\sum_{a\in A^j}v_a)$. In this function, it is possible to include all the costs that do not change with the number of travellers served and that the supplier has to bear in order to operate a mobility service. More specifically, these are related to investment costs, such as purchasing/leasing the fleet of vehicles, renting parking spaces, building charging stations, paying employees, and general legal and administrative costs for the company.
\begin{equation}
FC(v) = c_{lease}(\sum_{a\in A^j}v_a)
\label{Eq4}
\end{equation} 

The variable costs ($VC$) are associated to the daily operations, and they are defined as:
\begin{equation}
VC(\mathbf{f}) = \sum_{a\in A^{j}}(c_{a,fuel}l_{a})(1+f_{a})(1+\gamma)
\end{equation}
the costs $c_{a,fuel}$ incurred by the supplier when travellers use their service are mainly based on the distance travelled $l_a$. When considering sharing services, these factors have to be multiplied by the number of travellers using that mode of transport. Moreover, we include an additional component $\gamma$ connected with the relocation of vehicles or the return to a vehicle depot. 

Table \ref{table 2} shows the connection between the costs and revenues components of the profit maximization associated with the different modes of transport. The orange box indicates that a specific factor could influence the costs or revenues of an MSP based on their marketing strategies. The green box represents a component always considered for that specific mode of transport. When instead a component does not influence the profit of an MSP, empty boxes are shown in the table. 

 \begin{table}
 \tbl{Components of cost/revenues connected to modes of transport}
 {
 \begin{tabular}{p{0.15\textwidth}|p{0.1\textwidth}|p{0.1\textwidth}|p{0.1\textwidth}|p{0.1\textwidth}|p{0.1\textwidth}|p{0.1\textwidth}| p{0.1\textwidth}| p{0.1\textwidth}} 
 
 &\multicolumn{8}{c}{\textbf{Mode of Transport}} \tabularnewline \midrule
\textbf{Factor} & Bus & Train & Car-sharing one way& Car-sharing round trip & Bike-sharing one-way & E-scooter & Taxi & Carpooling 
\tabularnewline \midrule
 
 $c_{s}f_{a}$&\centering\cellcolor{lightsalmonpink}\CheckmarkBold   &\centering\cellcolor{lightsalmonpink}\CheckmarkBold  & \centering\cellcolor{lightsalmonpink}\CheckmarkBold &\centering\cellcolor{lightsalmonpink}\CheckmarkBold  & \centering\cellcolor{lightsalmonpink}\CheckmarkBold &\centering\cellcolor{lightsalmonpink}\CheckmarkBold && \tabularnewline \midrule
 
 $r_{s}f_{a}$&\centering\cellcolor{lightsalmonpink}\CheckmarkBold  &\centering\cellcolor{lightsalmonpink}\CheckmarkBold  &\centering\cellcolor{lightsalmonpink}\CheckmarkBold   &\centering\cellcolor{lightsalmonpink}\CheckmarkBold &\centering\cellcolor{lightsalmonpink}\CheckmarkBold  & \centering\cellcolor{lightsalmonpink}\CheckmarkBold &&\centering\cellcolor{lightsalmonpink}\CheckmarkBold \tabularnewline \midrule
 
 $c_{a,h}t_{a,main}(\textbf{f})f_{a}$&  &  &\centering\cellcolor{lightsalmonpink}\CheckmarkBold &\centering\cellcolor{lightsalmonpink}\CheckmarkBold & \centering\cellcolor{lightsalmonpink}\CheckmarkBold&\centering\cellcolor{lightsalmonpink}\CheckmarkBold&\centering\cellcolor{lightsalmonpink}\CheckmarkBold& \tabularnewline \midrule
 
 $c_{a,km}l_{a}f_{a}$& & \centering\cellcolor{lightsalmonpink}\CheckmarkBold &\centering\cellcolor{lightsalmonpink}\CheckmarkBold &\centering\cellcolor{lightsalmonpink}\CheckmarkBold&&&& \tabularnewline \midrule
 
 $c_{a,fixed}f_{a}$&\centering\cellcolor{lightsalmonpink}\CheckmarkBold &\centering\cellcolor{lightsalmonpink}\CheckmarkBold &\centering\cellcolor{lightsalmonpink}\CheckmarkBold &\centering\cellcolor{lightsalmonpink}\CheckmarkBold  &\centering\cellcolor{lightsalmonpink}\CheckmarkBold  &\centering\cellcolor{lightsalmonpink}\CheckmarkBold  & \centering\cellcolor{lightsalmonpink}\CheckmarkBold&\centering\cellcolor{lightsalmonpink}\CheckmarkBold \tabularnewline \midrule
 
$c_{lease}(\sum_{a\in A^j}v_a)$ &\centering\cellcolor{grannysmithapple}\CheckmarkBold  &\centering\cellcolor{grannysmithapple}\CheckmarkBold &\centering\cellcolor{grannysmithapple}\CheckmarkBold &\centering\cellcolor{grannysmithapple}\CheckmarkBold  &\centering\cellcolor{grannysmithapple}\CheckmarkBold  &\centering\cellcolor{grannysmithapple}\CheckmarkBold  &\centering\cellcolor{grannysmithapple}\CheckmarkBold&\centering\cellcolor{grannysmithapple}\CheckmarkBold  \tabularnewline \midrule

$c_{a,fuel}l_a$ &\centering\cellcolor{grannysmithapple}\CheckmarkBold &\centering\cellcolor{grannysmithapple}\CheckmarkBold & &  &  &  &&\centering\cellcolor{grannysmithapple}\CheckmarkBold  \tabularnewline \midrule

$c_{a,fuel}l_{a}f_{a}$ & & &\centering\cellcolor{grannysmithapple}\CheckmarkBold &\centering\cellcolor{grannysmithapple}\CheckmarkBold  &\centering\cellcolor{grannysmithapple}\CheckmarkBold  &\centering\cellcolor{grannysmithapple}\CheckmarkBold  & \centering\cellcolor{grannysmithapple}\CheckmarkBold& \tabularnewline \midrule

$c_{a,fuel}l_{a}f_{a}y$ &\centering\cellcolor{grannysmithapple}\CheckmarkBold &\centering\cellcolor{grannysmithapple}\CheckmarkBold &\centering\cellcolor{grannysmithapple}\CheckmarkBold &  &\centering\cellcolor{grannysmithapple}\CheckmarkBold  &\centering\cellcolor{grannysmithapple}\CheckmarkBold  &\centering\cellcolor{grannysmithapple}\CheckmarkBold & \tabularnewline
\bottomrule
\end{tabular}
}
\label{table 2}
\end{table}
 
Following Equation \ref{Eq1}, an MSP of the transport network will seek to maximise the profit, based on how many vehicles $v$ they operate and how they are distributed among the network links $v_a$. For this reason, the objective function of a MSP can be written in a compact form as:
\begin{equation}
    \max_{\mathbf{v_a}> \mathbf{0}} \quad  Pr(\mathbf{x^*}, v) = 
\sum_{s\in A^{j}_s}\sum_{a\in A^j} (C_{1}^{j}f_a + c_{a,h}t_{a,main}(\mathbf{f}) - c_{\text{lease}}(\sum_{a\in A^j}v_a) - C_{2}^{j})
\label{Eq 6}
\end{equation}
where the factor $C_{1}^{j}$ includes the component of costs associated with $f_a$ and $C_{2}^{j}$ the constant costs that affect the MSP profit.

\subsubsection{User groups and traffic assignment}
This section describes the traffic assignment model used in this paper. As anticipated in Section \ref{Sec2.1}, in order to represent the heterogeneity of users’ choices concerning their behaviours and travel perceptions \citep{book}, we divide the demand into $K$ classes. Each class of users is characterised by their sociodemographic characteristics, their home location and their daily trip chains, considered as a sequence of trips in a day. Users of the same class share the same origin, destination and all the locations visited in-between. At this stage, we do not explicitly consider the activities performed at each location; in future we could choose to define user classes so as to reflect components of the activity chains of users e.g. presence or absence of work activity.

Undoubtedly, the choice of dividing users into classes increases the complexity of the model. However, the socio-economic characteristics only affect users' perceived costs and not the network expansion. On the other hand, defining user classes based on the combinations of daily trip chains and activity locations  could become a non-trivial problem in large-scale networks. However, travel behavior literature shows that during typical weekdays the majority of travellers tend to perform home-work-home tours when using public transport or add an additional activity before/after work when travelling with private vehicles \citep{Sprumont2022QuantifyingTR,Axhausen2002ObservingTR}. Moreover, the combination of trip chains, activity sequences and locations are spatially limited and rather repetitive \citep{Susilo2014RepetitionsII}. Therefore, focusing on the most frequent tours, we cover most of the travel demand of an area and in turn keep the complexity of the supenetwork to a reasonable extent. 

As also mentioned in Section \ref{Sec2.1}, users’ trip chains are represented in the network as paths $p$ connecting OD pairs, and the traffic network conditions will be expressed in a path-based approach. The set of feasible paths encodes aspects of the multi-modal network, such as one-off subscription costs, and logical mode choice sequences. The computational expense and consequent limitations of path enumeration are well known. Although they are not problematic for the small examples presented here, future work will seek to tackle this issue. 

Based on the work of \cite{Nagurney2000AMM} on multi-class and multicriteria traffic network equilibrium, we write the relationship between link flow $f_a^k$ and path flow $x_p^k$ for class $k$ as:
\begin{equation}
f_a^k = \sum_{p \in P} x^{k}_p \delta_{ap}\quad\; \forall k\in K, \forall a\in A
\end{equation} 
Where the total flow on a link for a specific class is equal to the flow of that class on all the paths containing that link. Moreover, we could consider that the total flow on link $a$ depends on the sum of the link flow over all classes:
\begin{equation}
f_a = \sum_{k \in K} f^{k}_a \quad\; \forall a\in A
\end{equation} 
Furthermore, we consider a fixed demand model, in which the demand connecting each OD pair $w$ is known and constant in the considered time period. In this context, the sum of the path flows connecting $w$ for that for a class $k$ has to be equal to the total demand of that specific class $d_w^k$, as follows: 
\begin{equation}
d^{k}_w = \sum_{p \in P_w} x^{k}_p \quad \;\forall k\in K, \forall w\in W
\label{Eq 9}
\end{equation}

When making  their modal choices, users will encounter different costs associated with their chosen links (i.e. the mode-specific travel alternatives to reach the destination where the activity is performed). Each access link to a service will have a constant subscription cost or a fixed cost of ownership, if applicable to the mode of transport. Transfer links usually have zero cost when they are part of multi-modal paths; they can have the function of the access links and thus have a constant subscription cost ($c_s$). Two components of unitary costs characterise each mode-specific link: real monetary costs faced by the users to use the service associated with MSP, and class-dependent perceived costs connected to access time, waiting time, congestion, etc. The latter costs are flow and capacity-dependent functions. In some cases, the congestion effects can then be influenced by the demand flow of a specific mode of transport. In others, since each layer of the supernetwork represents the same underlying physical network, congestion can also be influenced by the flow of other modes of transport and by the flow of all classes of users crossing the same links. In particular, a class $k$ will perceive the cost on a generic mode-specific link $a$ of the network as follows:
\begin{equation}
C_a^{k}(\mathbf{f},v_a) = C_{a,access}^{k}(f_a,v_a) + C_{a,main}^{k}(\mathbf{f}) + C_{a,egress}^{k}(f_a,v_a)
\label{Eq 10}
\end{equation}
The first term of Equation \ref{Eq 10} represents the access cost that class $k$ will face in order to reach a mode of transport departing from an activity location. The access cost for user class $k$ is:

\begin{equation}
C_{a,access}^{k}(f_a,v_a) = c_{a,walking}^{k}t_{a,walking}(f_a,v_a) + c_{a,wait}^{k}t_{a,wait}(f_a,v_a)
\label{Eq 11}
\end{equation}

where the first component considers the time needed to reach the chosen mode of transport $t_{a,walking} (f_a  ,v_a )$. This function could be considered as a constant value, derived for example from the average distance from a bus stop, or it can be influenced by the users choosing the same mode of transport in relation to the limited capacity of the service. In this situation the users could choose to reach the next location in which to find more available vehicles, however at the cost of an increase in travel time. This time component is then associated with a monetary value of time, a cost that has a different weight based on the users’ class $c_{a,walking}^k$. The second component in Equation \ref{Eq 11} represents the time needed to wait for an available vehicle $t_{a,walking}(f_a,v_a)$, associated with the cost $c_{a,wait}^{k}$ perceived by class $k$. 

The second term in Equation \ref{Eq 10}, instead, represents the total cost for user class $k$ using the main mode of transport. Specifically, this cost can be calculated as follows:

\begin{equation}
C_{a,main}^{k}(\mathbf{f}) = c_{a,fuel}l_a + c_{a,km}l_a + c_{a,h}t_{a,main}(\mathbf{f}) + c_{a,ticket} + c^{k}_{a,main}t_{a,main}(\mathbf{f})
\end{equation}

The first term considers the cost for fuel (or electricity) based on the kilometres travelled. As explained in Section 2.1, the other three components are directly connected with the service chosen by travellers, and depends on the kilometres travelled, the time spent using the service, and an eventual fixed ticket or cost. The last term is connected to the cost $c_{a,main}^k$ associated to the time spent in the mobility service with the actual time $t_{a,main} (f)$, influenced by congestion effects. 

Finally, the last term presented in Equation \ref{Eq 10} indicates the egress cost of class $k$ on link $a$. 

\begin{equation}
C_{a,egress}^{k}(f_a,v_a) = c_{a,park}^{k}t_{a,park}(f_a,v_a) + c_{a,walking}^{k}t_{a,walking}(f_a,v_a)
\end{equation}

\begin{figure}
\centering
\includegraphics[width=\textwidth]{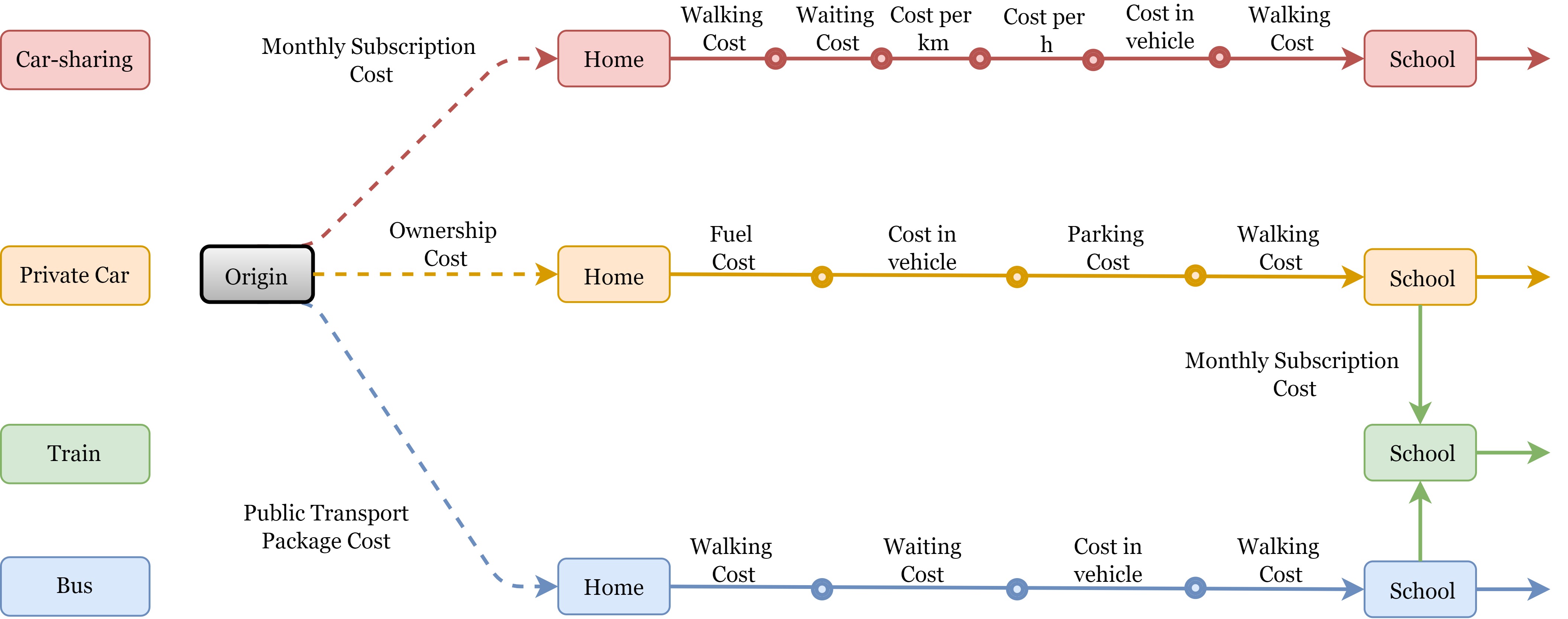}
\caption{Cost details of a link within the supernetwork shown in Figure \ref{Fig4}} \label{Fig5}
\end{figure}

These costs are calculated considering the time $t_{a,park} (f_a  ,v_a)$ spent to find a parking space with a specific mode of transport and the monetary cost $c_{a,park }^k$ that class $k$ associates with this time. In addition, the cost of the time needed to reach the final destination is considered, once left the main mode of transport, multiplied by the time.
Figure \ref{Fig5} details a section of the supernetwork of Figure \ref{Fig4}, to illustrate how these costs are assigned based on the mode of transport considered.

Following this approach, the supernetwork is built in order to maintain link-additive costs. Therefore, the cost for user class $k$ of a path connecting an OD pair is equal to the sum of its constituent link costs:
\begin{equation}
C_{p}^{k}(\mathbf{x},v) = \sum_{s\in A_s}\sum_{a\in A}c_s\delta_{a,s}\delta_{a,p} + \sum_{a\in A} C^{k}_{a}(\mathbf{f},v_a)\delta_{a,p}
\end{equation}
Once the supernetwork is constructed and link costs defined, all users of each class are assigned to the network following Wardop’s first principle \citep{Wardrop1952SomeTA}. For each class, for all OD pairs and for all paths, the path flow vector $\textbf{x}^*$ is said to be an equilibrium if the following conditions hold:
\begin{equation}
    C_{p}^{k}(\mathbf{x}^*,v) \begin{cases} = \rho^{k}_w & x_{p}^*>0 \\ \geq \rho^{k}_w & x_{p}^*=0 \end{cases}
    \label{Eq 15}
\end{equation}
with
\begin{equation}
    x_{p}\geq0 \quad \; \forall p\in P
    \label{Eq 16}
\end{equation}
where (\ref{Eq 16}) is the path flow non-negativity constraint. The described model is based on non-separable link cost functions, due to the fact that the cost on a link is influenced by the flow on other links of the supernetwork.
Following \cite{Nagurney1992NetworkEA}, $\textbf{x}^* \in \Phi$ (a nonempty closed convex subset of $R^n$) satisfies the equilibrium conditions (\ref{Eq 15}) if it satisfies the VI problem:
\begin{equation}
    \sum_{k\in K}\sum_{w\in W}\sum_{p\in P_w}C^{k}_{p}(\mathbf{x}, v)(\xi_{p} -x_p)\geq0 \quad \;  \forall \xi_{p} \in \Phi
    \label{Eq17}
\end{equation}

We denote by $\Phi$ the set of all demand feasible non-negative path flows (i.e. satisfying Equation \ref{Eq 9}).
\cite{Nagurney2000AMM} demonstrates, as a qualitative property of the formulation, the existence of a solution of the VI (\ref{Eq17}); nevertheless, in this case we cannot prove uniqueness of solutions, due to the non-separability of the link costs. In the special case of separable transport modes and hence link costs, uniqueness of equilibrium can be guaranteed. An example could be a network in which the trips can be done with private car on the road, a bus service and a train service on a dedicated and separated infrastructure system. 

\section{Mathematical Program with Equilibrium Constraints (MPEC)} \label{sec3}
In order to study the interaction between an MSP, responding to a profit maximization objective function (Eq. \ref{Eq 6}), and multi-class users, finding their network equilibrium in the form of a VI (Eq. \ref{Eq17}), we formulate the problem as an MPEC. Such a program is a generalisation of a bi-level program, where the inner program is defined through VIs \citep{Luo1996MathematicalPW}. 
The problem can then be written in a compact form as follows:
\begin{equation}
    \max_{\mathbf{v\in \mathbb{V}}} \quad  
Pr(\mathbf{x^*},v)
\end{equation}
\begin{align} 
\textrm{s.t.} \quad & \mathbf{x^*} \in \mathbb{X}^*(v)
\end{align}
where $\mathbb{X}^*(v)$ is a subset of $\mathbb{X}$ that satisfies Equation \ref{Eq17}. It is important to underline that the supplier’s objective function contains the lower-level variables, and as a consequence the lower-level equilibrium strictly depends on the number of vehicles available for each trip.

\subsection{Solution algorithm} \label{sec 3.1}
Several iterative methods have been proposed to solve MPECs, such as penalty interior point approach (PIPA), Implicit Programming Algorithm (IMPA), and Newton type approaches \citep{Luo1996MathematicalPW, phdthesis}. In this paper, due to the non-separability of the problem, an iterative solution algorithm is adopted. We search for a local solution of the upper level continuous objective function using a gradient-based method, based on the sequential quadratic programming (SQP) algorithm \citep{wright1999numerical}. We have implemented this approach using the conventional MATLAB function \texttt{fmincon} with multiple variables, including its optional settings. The lower-level equilibrium is, instead, calculated with the MPM (or Extragradient Method). This algorithm, proposed to solve the multi-modal traffic assignment VI \citep{Nagurney2000AMM,NAGURNEY2002445, SZETO201451}, divides the VI problem into a series of quadratic programming problems in which the constraints are linear \citep{10.2307/2345190}. The initial objective function with its constraints takes the form:

\begin{equation}
    \min \frac{1}{2}x^{T}Qx-bx
    \label{Eq 20}
\end{equation}
subject to $Ax\leq z$ and $x\geq0$ and where Q is a positive semidefinite matrix. 

In order to apply the MPM, let’s consider M as a nonempty closed convex subset of $R^n$, and a continuous function $C_p^k (\mathbf{x^*},v)$, $\mathbf{x^*}$ is a solution vector of the user equilibrium if it satisfies the VI problem of Equation \ref{Eq17}. Algorithm \ref{Al1} describes the steps of the MPM, starting with an initial value of $\mathbf{x^*}$ for iteration $i$, then it updates $\mathbf{x}$ at each iteration $i$ calculating a first value of  $\overline{\rm \mathbf{x}}^{i}$ as follows:
\begin{equation}
    \overline{\rm \mathbf{x}}^{i} = P_{M}(\mathbf{x}^{i} - \psi C^{k}_{p}(\mathbf{x}^{i},v))
    \label{Eq 21}
\end{equation}
with $\psi$ is a constant positive step length between the values 0 and 1. When the value of $\psi$ is too small we have a slow convergence, in case of a big value of $\psi$, instead, it can be hard to achieve convergence. In Equation \ref{Eq 21}, $P_M(\bullet)$ is the orthogonal projection onto M and a solution of the quadratic problem:
\begin{equation}
    \min \frac{1}{2}\mathbf{x}^{T}\mathbf{x}-(\mathbf{x}^{i} - \psi C^{k}_{p}(\mathbf{x}^{i},v))^{T}\mathbf{x}
\end{equation}

\begin{algorithm}[H]
\caption{Modified Projection Method}
\label{Al1}
\begin{algorithmic}[1]
\STATE \textbf{Input} Parameters and Functions: $J,KS,A,N,P_{w}, d^{k}_{w},c_{s},c_{a},c^{k}_{a},\delta, l_{a}, \mathbf{x}^{i},v$
\STATE \textbf{Initialization} Set $MaxIterations, \psi, \chi^1, \chi^2, i = 1, \mathbf{x}^{i}$
\STATE \textbf{Calculate} $b = \mathbf{x}^{i} - \psi C(\mathbf{x}^{i},v)$ 
\STATE \textbf{Solve} $\overline{\rm \mathbf{x}}^{i} = \min_{x} \frac{1}{2}\mathbf{x}^{T}\mathbf{x}-b\mathbf{x}$ 
\STATE \textbf{Calculate} $b = \mathbf{x}^{i} - \psi C(\overline{\rm \mathbf{x}}^{i},v)$ 
\STATE \textbf{Solve} $ \mathbf{x}^{i+1} = \min_{x} \frac{1}{2}\mathbf{x}^{T}\mathbf{x}-b\mathbf{x}$ 
\STATE \textbf{Calculate} $gap$ 
\IF{$|\mathbf{x}^{i+1} - \mathbf{x}^{i}| 	\geq \chi^2 \land gap \geq \chi^1$}
\STATE Set $\mathbf{x}^{i} = \mathbf{x}^{i+1}$
\STATE $i = i+1$ 
\IF{$i \geq MaxIterations$}
\STATE \textbf{Stop}
\ELSE
\STATE Go to 3
\ENDIF
\ELSE
\STATE Set $\mathbf{x}^{*} = \mathbf{x}^{i+1}$
\ENDIF
\end{algorithmic}
\end{algorithm}

In this reformulation of Equation \ref{Eq 20} the symmetric matrix Q becomes the identity matrix, and $b$ is the value in brackets. The solution of the problem is calculated considering as constraints of the problem the equality constraint \ref{Eq 9} and the path flow nonnegativity constraint \ref{Eq 16}.
Once the value of $\overline{\rm \mathbf{x}}^i$ is found solving Equation \ref{Eq 21}, a second projection is necessary:
\begin{equation}
    x^i = P_{M}(\mathbf{x}^{i} - \psi C^{k}_{p}(\overline{\rm \mathbf{x}}^{i},v))
\end{equation}
as a solution of the quadratic problem.

At the end of each iteration, a relative gap function \citep{Chiu2011DynamicTA} is calculated in order to verify the convergence of the model:
\begin{equation}
    gap = \sum_{k \in K} \sum_{w \in W} \frac{[\sum_{p \in P_{w}}(x^{k}_{p}* C^{k}_{p}(\mathbf{x},v))] - d^{k}_{w}*C^{k}_{p_{w}min}(\mathbf{x},v)}{d^{k}_{w}*C^{k}_{p_{w}min}(\mathbf{x},v)} < \chi^1
\end{equation}
Simultaneously, an additional convergence verification is applied $|\mathbf{x}^{i+1}-\mathbf{x}^i |<\chi^2$. When convergence is verified, an equilibrium solution of the vector of path flows $\mathbf{x}^*$ is deemed to have been found. Naturally, the modeller can define different thresholds. Through this value it is possible to calculate the value of the profit function and apply the upper-level optimization procedure. 

\section{Numerical Examples}

In this section we present three examples from which we will illustrate the key properties of the model formulated in this study. First, we introduce an example to show the different characteristics of the methodological approach, including a demonstration of the importance of the non-separability assumption included in the model. Secondly, we show how to encode in the network a mobility package, in order to analyse the influence of different pricing strategies to maximize a supplier's profit. Finally, by introducing a new competitor in the network, we show how different strategies can lead to equilibria where a MSP can attract or not demand inside the transportation market. 

\subsection{Example 1: A single MSP} \label{sectex1}
\begin{figure}[ht]
	\centering
	\subfloat[Daily tours]{%
		\includegraphics[width=.7\columnwidth,keepaspectratio]{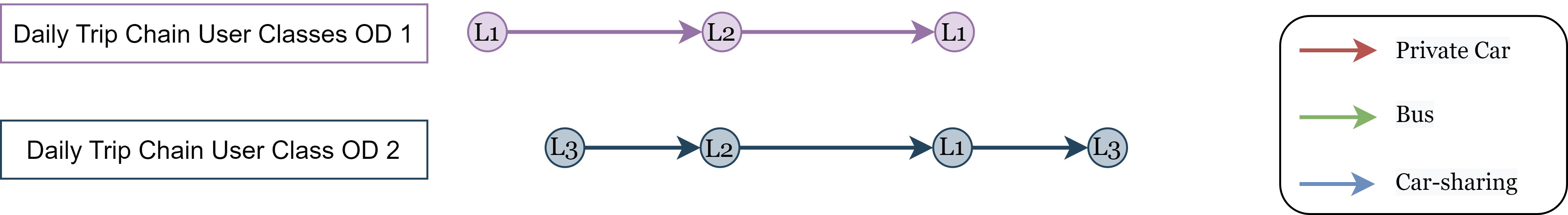}
  \label{Fig6a}}%
 \quad%
	\subfloat[Supernetwork expansion]{%
		\includegraphics[width=.7\columnwidth,keepaspectratio]{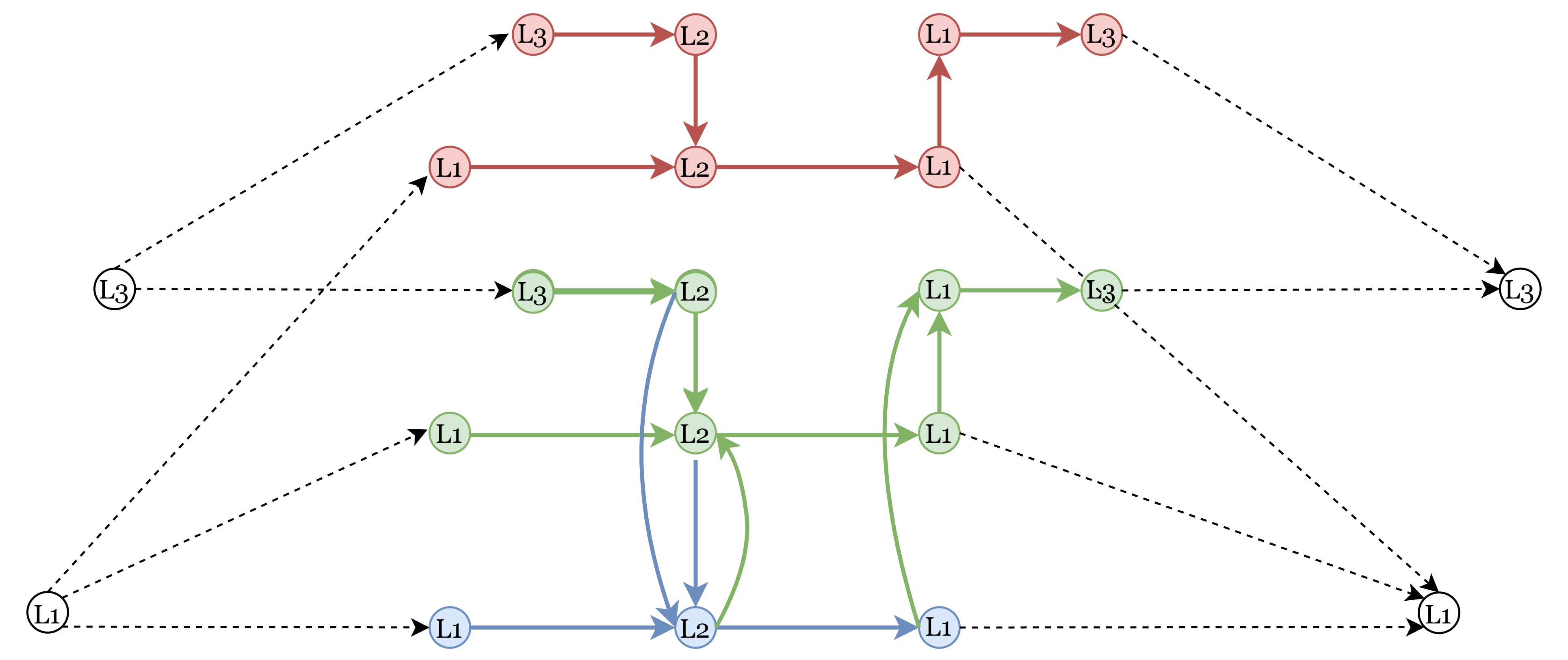}
            \label{Fig6b}%
	}%
    \caption{Example 1}
	\label{Figure 6}%
\end{figure}

In a first example we define a small sized problem (Figure \ref{Figure 6}) involving three user classes that are performing distinct daily tours in different locations (L), shown in Figure \ref{Fig6a}. Due to their socio-economic characteristics, in the first daily trip chain (represented by purple nodes), two classes of users will perceive travel costs differently (specified in table \ref{table 3} as class 1 and 2). An additional class (identified as class 3), instead, is performing a different daily tour illustrated with dark blue nodes in Figure \ref{Fig6b}. The demand 
for the three classes is assumed respectively 200, 100 and 200. 

The available modes of transport considered in the example network are: private car, bus, and one-way car-sharing service. Intentionally, the different classes of users share a link connection, between location L2 and location L1. Congestion effects will affect users that choose car and car-sharing modes, since they are assumed to share the same links on the base network. The bus service, instead, is considered to have a dedicated lane. In this case, the attractiveness of the service is mainly affected by the waiting time at the bus stop. We calculate the MPEC focusing on the profit maximization goal sought by the car-sharing service in the different scenarios; nevertheless the same analysis can be carried out considering other suppliers at the upper level. Table \ref{table 3} lists all the parameters used to calculate the equilibrium of the multi-modal supernetwork, and table \ref{table 4} shows the functional parameters. Elements not listed were not considered relevant for the specific mobility services taken into account. In this example, the link costs functions take the form of the conventional Bureau of Public Roads (BPR) function (Equation \ref{Eq 25}), or they are considered constant. 

\begin{equation}
    t = t_{0}\left(1 +\alpha\left(\frac{f}{C}\right)^{\beta}\right)
    \label{Eq 25}
\end{equation}

\begin{table}[ht]
\tbl{Parameters Example}
{
\begin{tabular}{c|c|c|c} 
\toprule
 \textbf{Parameters} & \textbf{Private Car} &\textbf{Bus} & \textbf{Car-sharing}\\ \midrule
$c_s$  & -  & 1.5 & 2\\ 
$c^{1}_{a, access}/c^{1}_{a, egress}$  &8 &8.2& 8\\
$c^{2}_{a, access}/c^{2}_{a, egress}$  & 9.6& 9.8 & 9.6\\
$c^{3}_{a, access}/c^{3}_{a, egress}$  & 8& 8.2 & 8\\
$c^{1}_{a,wait}$  &- &10 & 8\\
$c^{2}_{a,wait}$   &-&12&9.6 \\
$c^{3}_{a,wait}$   &-&10& 8\\
$c^{1}_{a,main}$  &8 &10 &8\\
$c^{2}_{a,main}$   &9.6 &9.8 &9.6\\
$c^{3}_{a,main}$   &8 &10 &8\\
$c_{a, fuel}$  &0.37 &0.05 &0.3\\ 
$c_{a,h}$  &- &- &0.6 \\
$c_{a,km}$  &- &-&0.6\\
$l_{a}$  &10 &10 &10 \\
$c^{1}_{a,park}$  &9 &- &- \\
$c^{2}_{a,park}$  &10.8 &- &- \\ 
$c^{3}_{a,park}$  &9 &- &- \\ 
 \bottomrule

\end{tabular}}
\label{table 3}
\end{table}

\begin{table}[ht]
\tbl{Functional Parameters}
{
\begin{tabular}{c|c|c|c|c|c|c|c|c|c|c|c|c|c|c|c|c} 
\toprule
\textbf{Function} & \multicolumn{5}{c|}{\textbf{Private Car}}  &\multicolumn{5}{c|}{\textbf{Bus}} & \multicolumn{5}{c}{\textbf{Car-sharing}}\\ \midrule
 &$t_0$&$\alpha$&$f$& $C$ &$\beta$&$t_0$&$\alpha$&$f$& $C$ &$\beta$&$t_0$&$\alpha$&$f$& $C$ &$\beta$\\\cline{2-16}
$t_{a, access}(f_{a},v_{a})$&\multicolumn{5}{c|}{-} &0.0625&\multicolumn{4}{c|}{-}&0.0125& 0.15& $f_a$&$v_a$& 4 \\ \midrule
$t_{a,wait}(f_{a}, v_{a})$&\multicolumn{5}{c|}{-}   &0.3& 0.15&$f_a$ &150&2&0.05& 0.2&$f_a$&$v_a$& 4&\\ \midrule
$t_{a,main}(\textbf{f})$&0.2&4&$\textbf{f}$ &250& 2& 0.4&\multicolumn{4}{c|}{-}&0.2&4&$\textbf{f}$ &250& 2&\\\midrule
$t_{a,park}(f_{a}, v_{a})$&0.1& 2.5&$f_a$ &250 &2&\multicolumn{5}{c|}{-} &\multicolumn{5}{c|}{-} \\ \midrule
$t_{a, egress}(f_{a}, v_{a})$ &0.075&\multicolumn{4}{c|}{-} &0.0625&\multicolumn{4}{c|}{-} &0.0125& 0.15& $f_a$&$v_a$& 4\\\midrule
$c_{lease}(v)$ &\multicolumn{5}{c|}{-}&\multicolumn{5}{c|}{$6v$}&\multicolumn{5}{c}{-} \\
 \bottomrule
\end{tabular}}
\label{table 4}
\end{table}

\begin{figure}[ht]
	\centering
	\subfloat[Upper level solution]{%
		\includegraphics[width=.47\columnwidth,keepaspectratio]{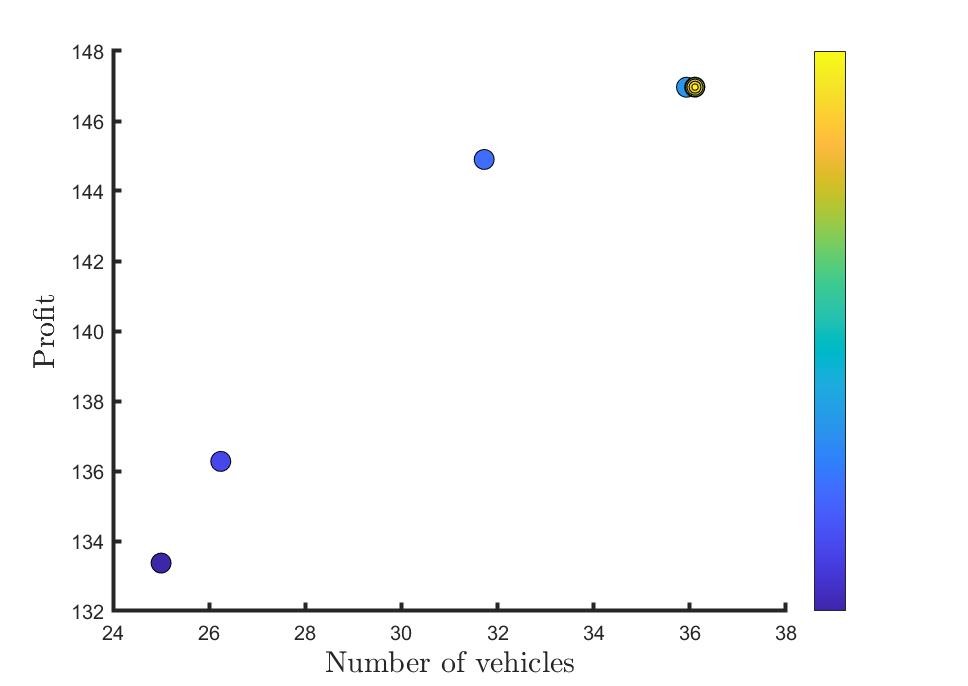}
  \label{Fig7a}}%
 \quad%
	\subfloat[Lower level Relative Gap]{%
		\includegraphics[width=.47\columnwidth,keepaspectratio]{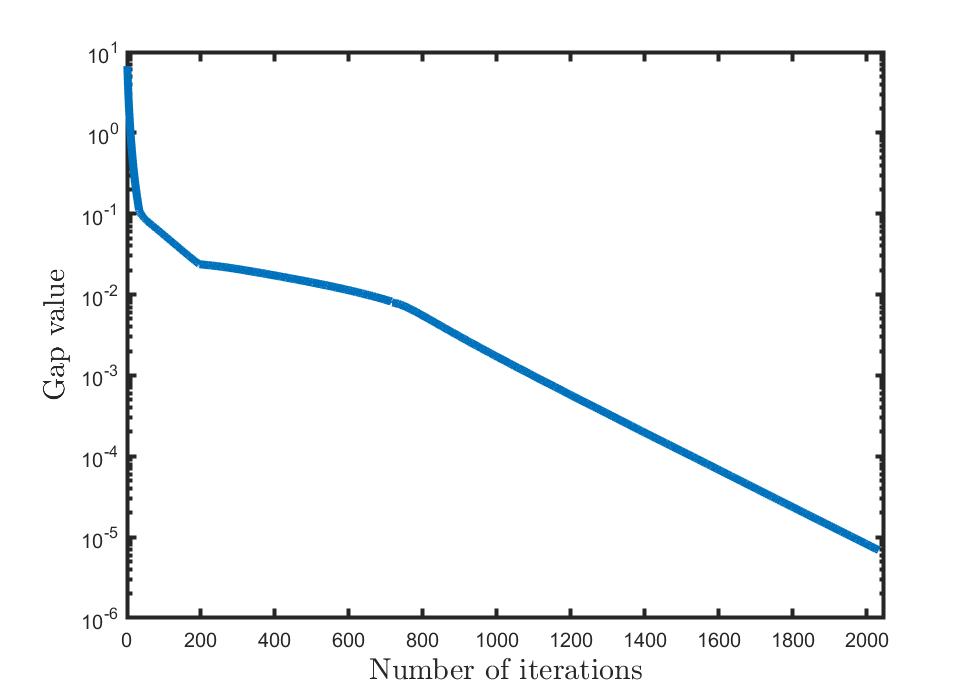}
            \label{Fig7b}%
	}%
    \caption{MPEC solution}
	\label{Fig7}%
\end{figure}

We apply the solution algorithm described in Section \ref{sec3} using MATLAB R2019b\footnote{The simulations are carried out  using Windows 10 laptop with an Intel(R) Core (TM) i7-8650U CPU with a base frequency of 1.90GHz and a system memory of 16.0 GB.}, imposing a value of $\psi =0.5$ for the MPM. The optimal solution of the MPEC is calculated in 446.2 seconds, realizing 9 iterations for the upper level objective function as shown in Figure \ref{Fig7a}. In the Figure, the first iteration corresponds to the dark blue colour and the solution is characterized by the yellow colour.  Figure \ref{Fig7b} shows, instead, the relative gap variation computing 2033 iterations at the lower level for the optimal solution of the upper level.

\begin{figure}[ht]
	\centering
	\subfloat[Bus Costs Variation]{%
		\includegraphics[width=.4\columnwidth,keepaspectratio]{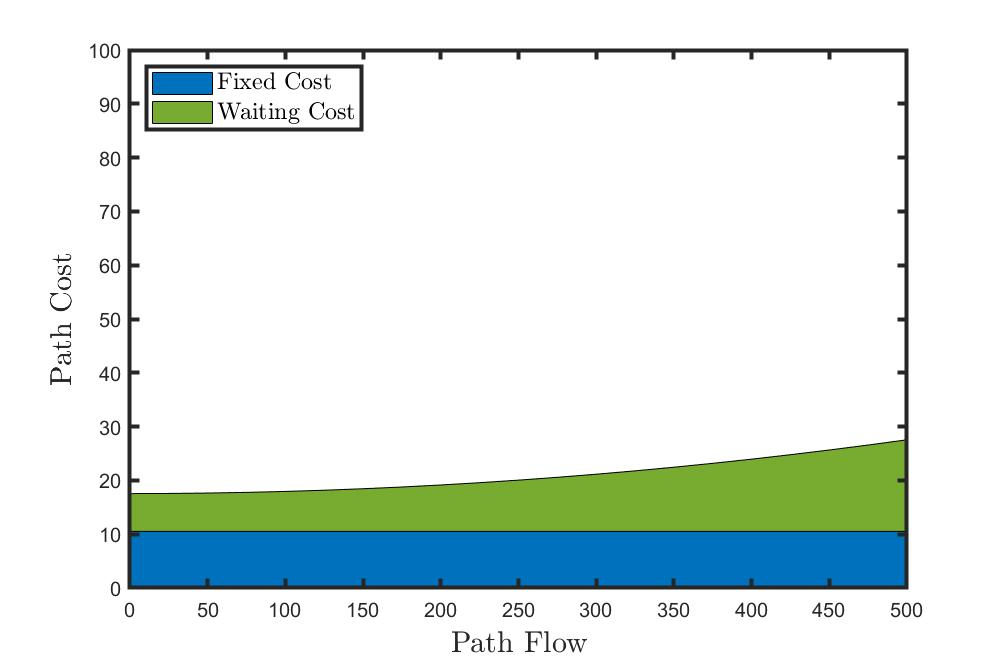}
  \label{Fig8a}}%
 \quad%
	\subfloat[Private Car Costs variation]{%
		\includegraphics[width=.4\columnwidth,keepaspectratio]{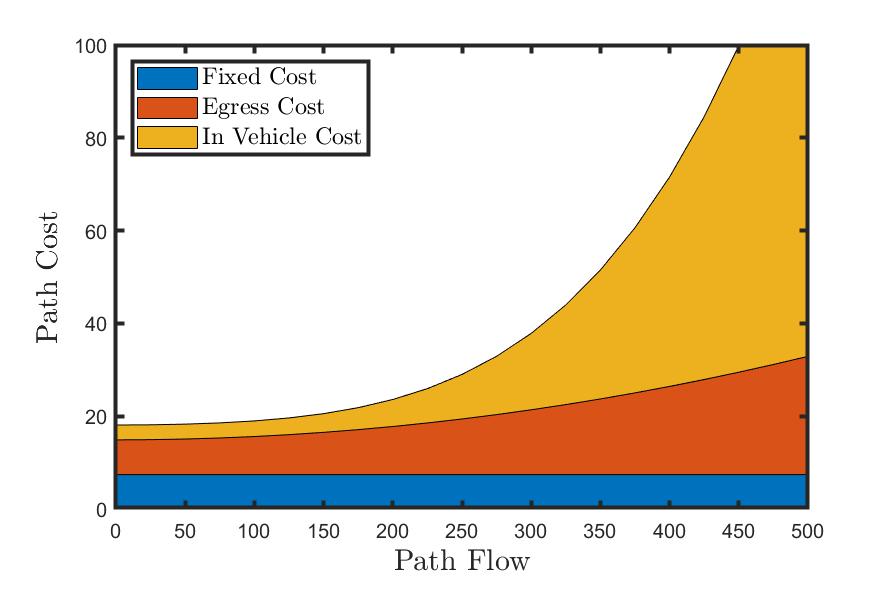}
            \label{Fig8b}%
            
	}%
  \quad%
	\subfloat[Car-sharing Costs variation]{%
		\includegraphics[width=.4\columnwidth,keepaspectratio]{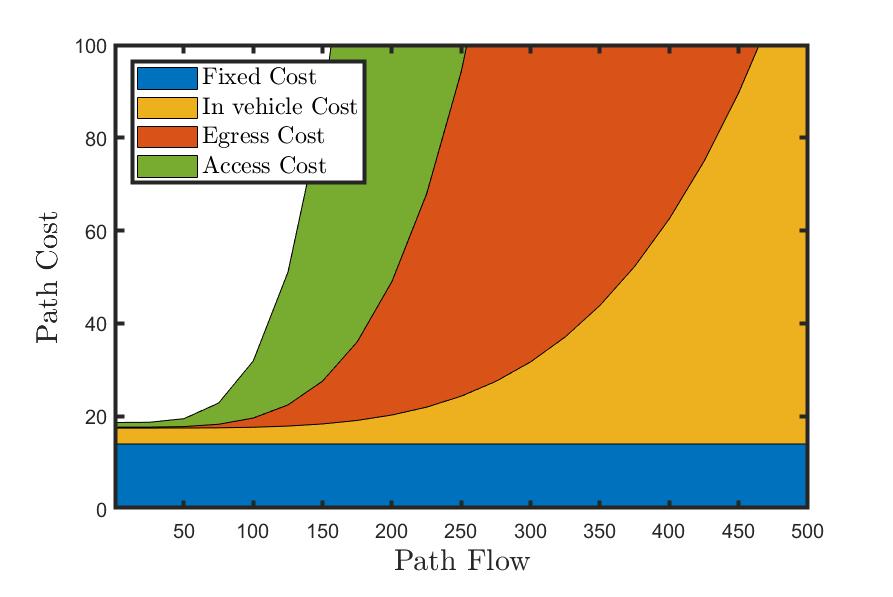}
            \label{Fig8c}%
            
	}%
    \caption{Path Costs variations}
	\label{Fig8}%
\end{figure}

In Figure \ref{Fig8}, we show how the different cost components are influencing each mode of transport with the variation of the path flow. We evaluated these values imposing a fixed number of vehicles for the car-sharing service equal to the output of the MPEC solution ($v = 36$). The bus cost variation (Figure \ref{Fig8a}) clearly has a strong constant component, due to the assumptions made for this mode of transport. Different behaviour can be instead observed for car and the car-sharing service. The private car cost (Figure \ref{Fig8b}) is mainly influenced by the congestion on the road, while the time spent to find an available parking spot or to walk to reach the final destination slightly increases only when the path flow reaches high values. On the contrary, the car-sharing cost is mainly affected by the access and egress costs. The growth in travel demand is therefore converted in an increase of the walking distance to reach a station or waiting time to find an available vehicle. 

\begin{figure}[ht]
\centering
\includegraphics[scale=0.3]{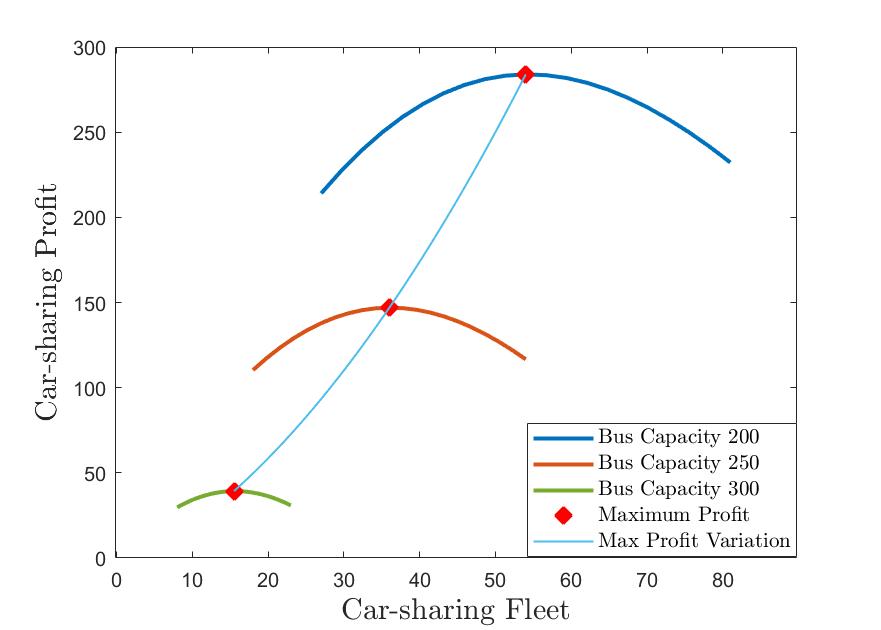}
\caption{Car-sharing profit variation with Bus capacity}
\label{Fig22}
\end{figure}

Finally, in Figure \ref{Fig22}, we evaluate how the car-sharing profit varies when another MSP present in the network decides to change their service capacity. Specifically, the light blue curve shows how the car-sharing maximum profit increases when the bus operator decreases their capacity, and vice versa.

\subsubsection{Methodological insight: the relevance of non-separability of the cost functions}
\begin{figure}[ht]
\centering
\includegraphics[scale=0.3]{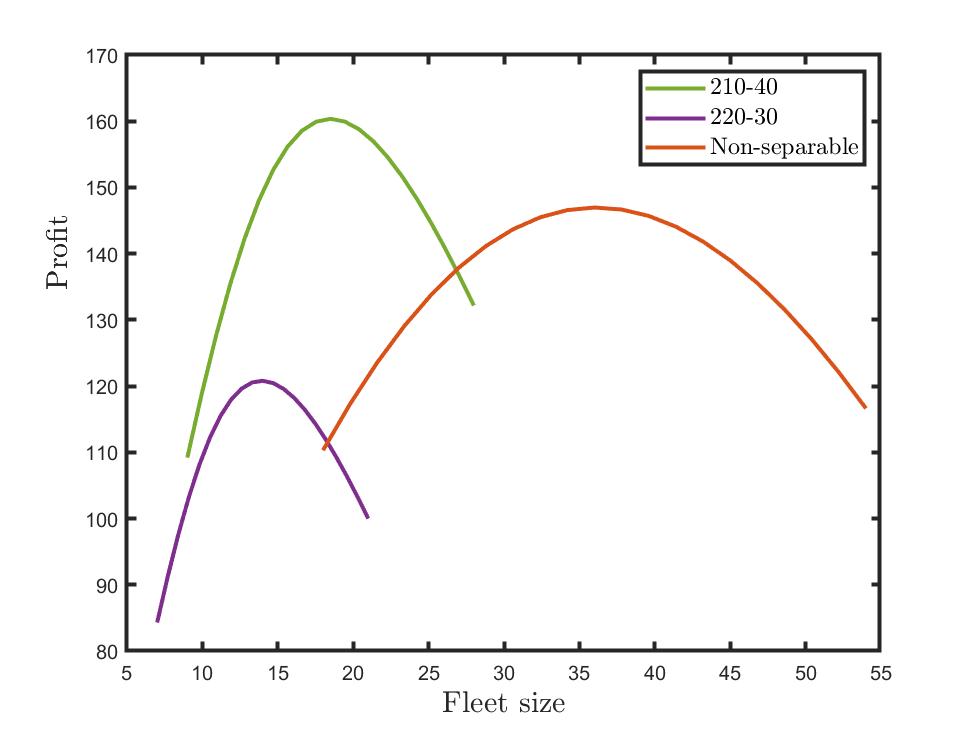}
\caption{Car-sharing profit variation with capacity split}
\label{Figure 9.1}
\end{figure}

In this additional section, we would like to show why the non-separability assumption made on link cost functions is fundamental to develop a more realistic model. In order to do so, we apply the solution algorithm introduced in section \ref{sec 3.1} to the network illustrated in Figure \ref{Figure 6}, using the parameters and functions introduced in Table \ref{table 3} and \ref{table 4}, assuming that the different modes of transport are characterised by separable cost functions. Although we relax our assumptions, private car and the car-sharing service are still part of the same infrastructure, and how the available road capacity is split between the two services has an important impact of the equilibrium solution as well as on the provider's profit. For this reason, we evaluate how the profit curve and the fleet size vary based on capacity split. To underline the importance of the presence of non-separable cost functions in these type of problems, in Figure \ref{Figure 9.1} we compare this scenario with the base example previously introduced.

In Figure \ref{Figure 9.1} we illustrate only those profit curves (from all possible capacity splits) that are comparable to the non-separable case, associated to value of capacity split [220-30] and [210-40], where the first value indicates the capacity of the private car  infrastructure and the second the car-sharing infrastructure. The red curve, instead, represents the non-separable case. 

From this analysis, it is clear that neglecting the influence that different modes of transport have on each other could lead to an incorrect estimation of the profit, and more importantly to an underestimation of the number of vehicles needed to offer a profitable service. 

\subsection{Example 2: Cooperation between MSPs via mobility package}
\begin{figure}[!tb]
	\centering
	\subfloat[Daily tours]{%
		\includegraphics[width=.7\columnwidth,keepaspectratio]{Figures/Fig6_1.jpg}
  \label{Fig12a}}%
 \quad%
	\subfloat[Supernetwork expansion]{%
		\includegraphics[width=.7\columnwidth,keepaspectratio]{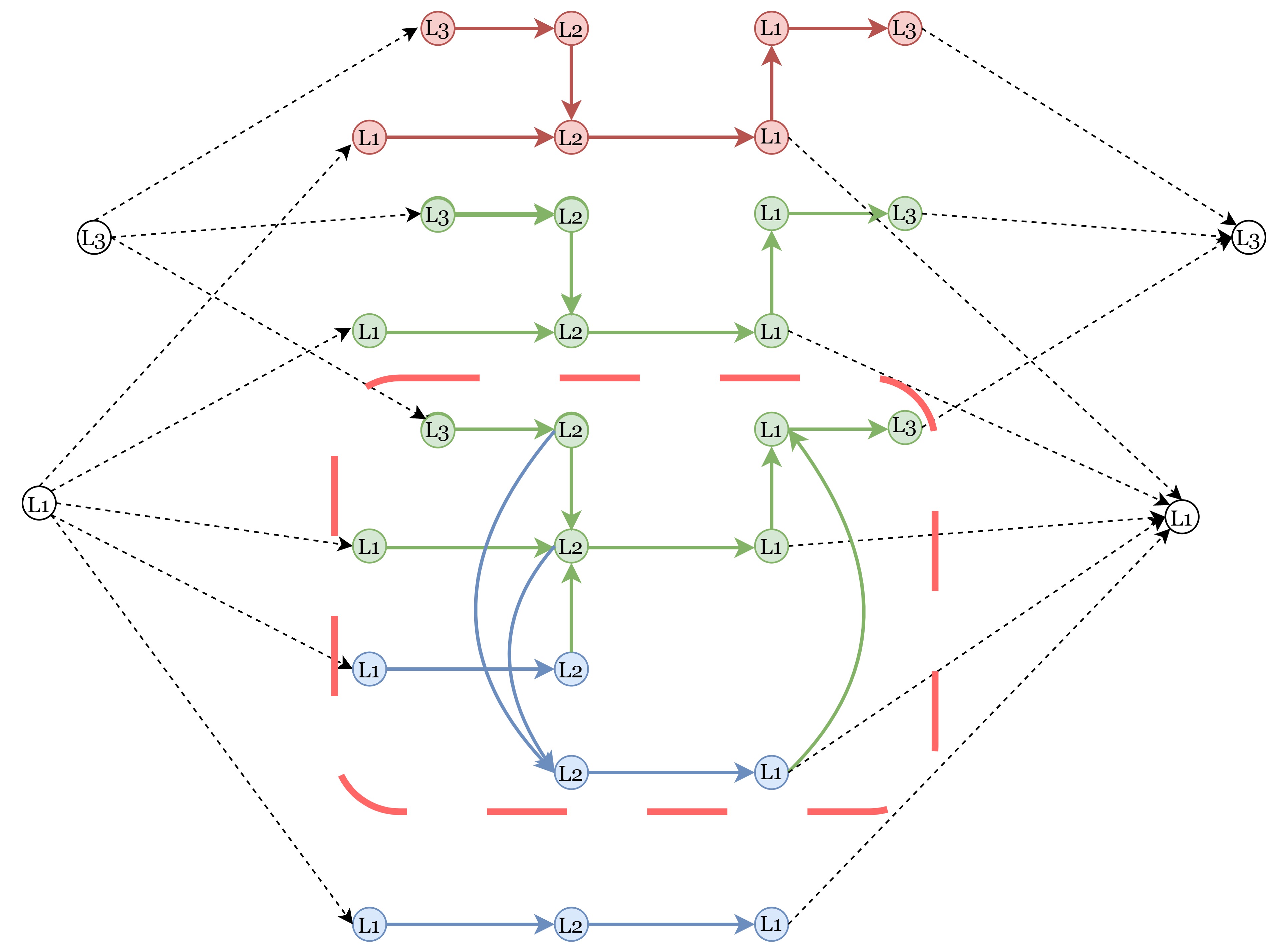}
            \label{Fig12b}%
	}%
    \caption{Example 2 with package Bus + Car-sharing}
	\label{Figure 9}%
\end{figure}

From the network introduced in Section \ref{sectex1}, in this second example, we consider that the bus provider and the car-sharing supplier decide to collaborate selling an integrated mobility package, identified by the sub-graph inside the red-dashed bounding box in Figure \ref{Figure 9}, through which travellers can pay an attractive price to use both services.

 The network is built so that users can subscribe to the monthly package to use the combination of bus and car-sharing, or bus with a reduced price compared to the pay-as-you-go option. Users that buy the mobility package have to make at least one of their trips by bus, otherwise they are not allowed to use it. Alternatively to the new package, users can still travel by bus buying a daily ticket, use their private car, or subscribe to a monthly package using the car-sharing service. Hence, the services are still offered as separated, but the integrated package is straightforwardly assumed to be more convenient than the separated offers.
 
The parameters used in the example are the same introduced in Table \ref{table 3} and \ref{table 4}. For the car-sharing service, we considered a fixed capacity equal to the optimal value obtained on Example 1, and we apply the MPEC to evaluate the additional fleet size necessary after the introduction of the collaboration with the bus operator. 
\begin{figure}[h!]
\centering
\includegraphics[scale=0.35]{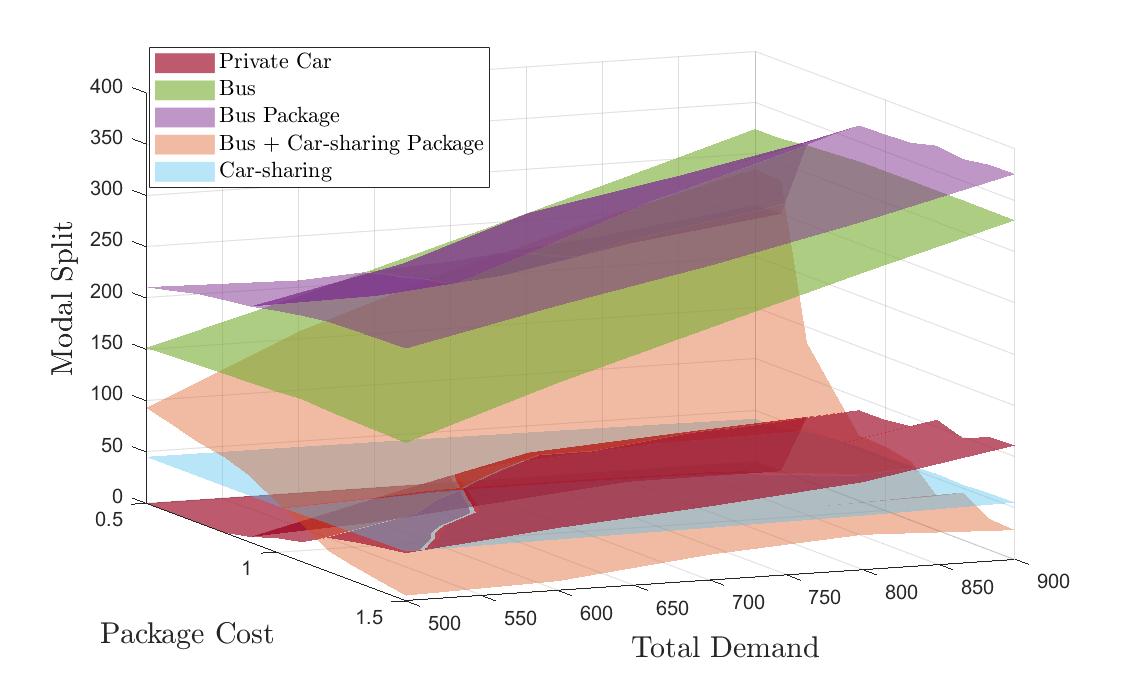}
\caption{Modal split variation with package cost}
\label{Figure 13}
\end{figure}
The structure of the proposed model allows us to study different pricing strategies that can be tested to evaluate the system equilibrium. In this scenario, specifically, we want to understand how much the price of the package can influence the usage of the car-sharing service and the profit of the car-sharing supplier. 

In this analysis we consider the package price increasing from 0.5 to 1.5 € per day. The revenue for the bus provider is considered to be fixed and equal to 0.5 € per day per traveller; the car-sharing revenue, instead, varies in between 0 and 1 € per day per traveller. We considered that the car-sharing supplier will get the package revenues based on the number of subscribers using their service. The results are shown in Figure \ref{Figure 13}, where the 3D plot illustrates the change of modal split based on variation of the package price, while varying the total demand between 500 and 900, keeping the same class division.  

\begin{figure}[p]
  \centering
  \subfloat[Total Demand: 500]{
  \includegraphics[width=.45\textwidth]{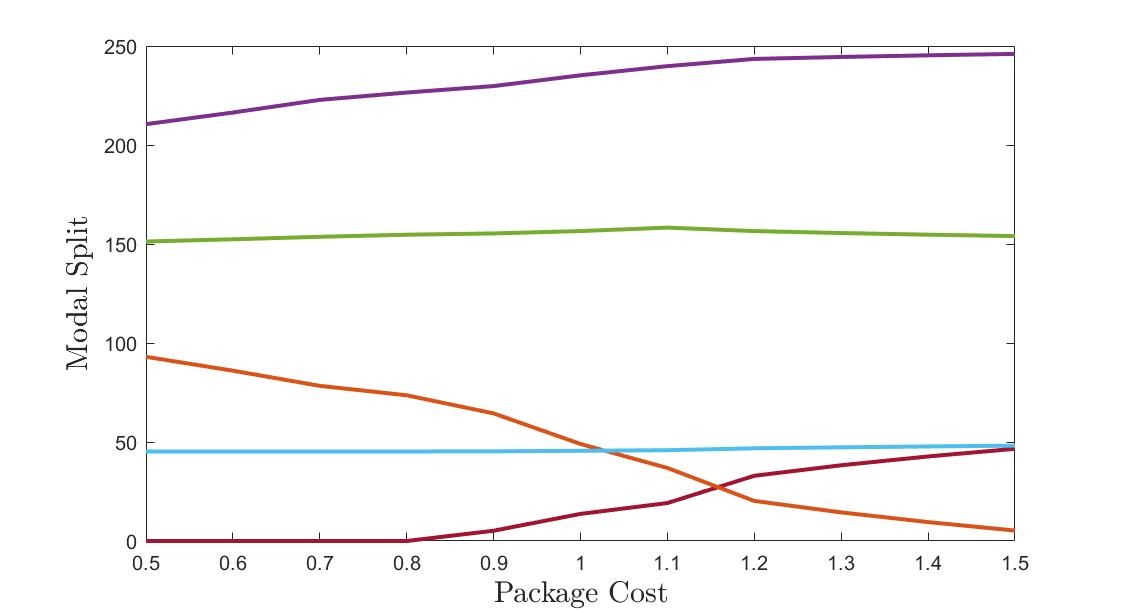}}
  \hspace{0.5cm}
  \subfloat[Package Cost: 0.5]{
  \includegraphics[width=.45\textwidth]{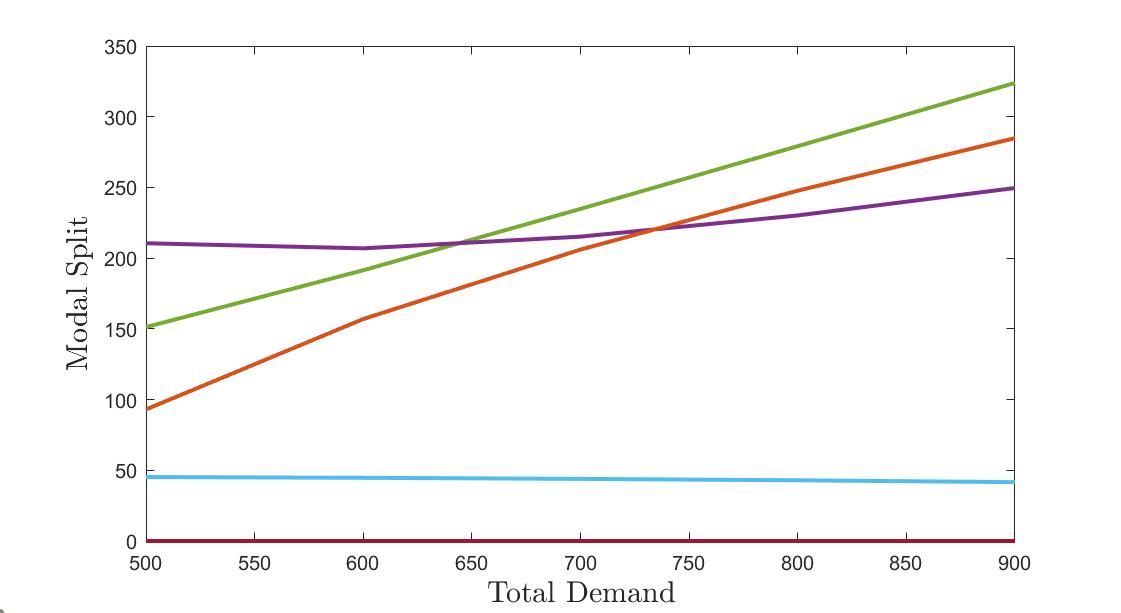}}

  \vspace{0.5cm}
\subfloat[Total Demand: 700]{
  \includegraphics[width=.45\textwidth]{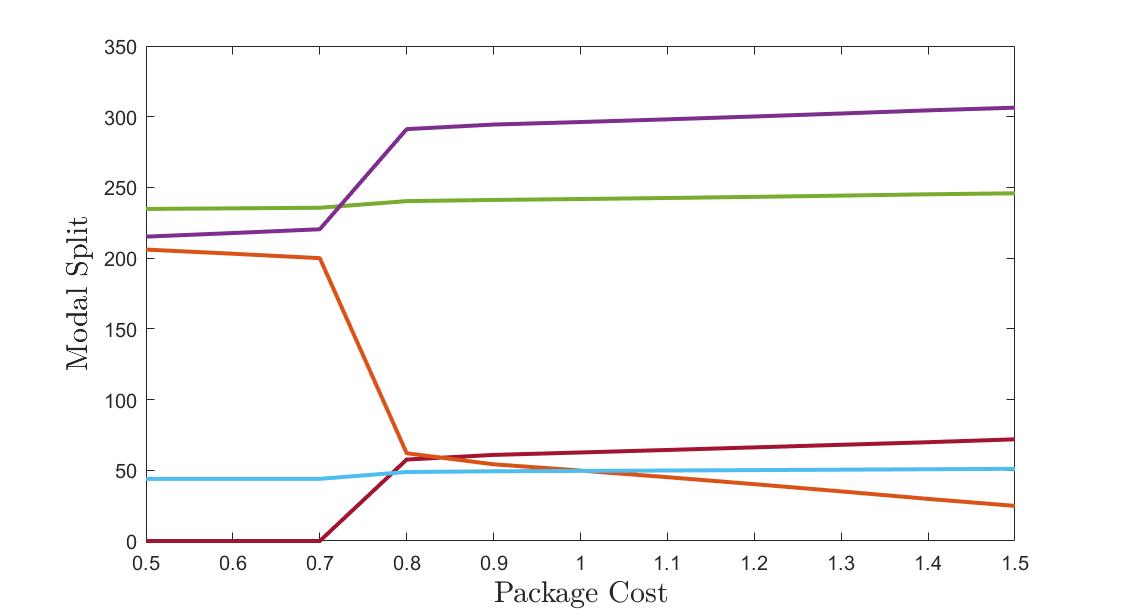}}
  \hspace{0.5cm}
  \subfloat[Package Cost: 0.9]{
  \includegraphics[width=.45\textwidth]{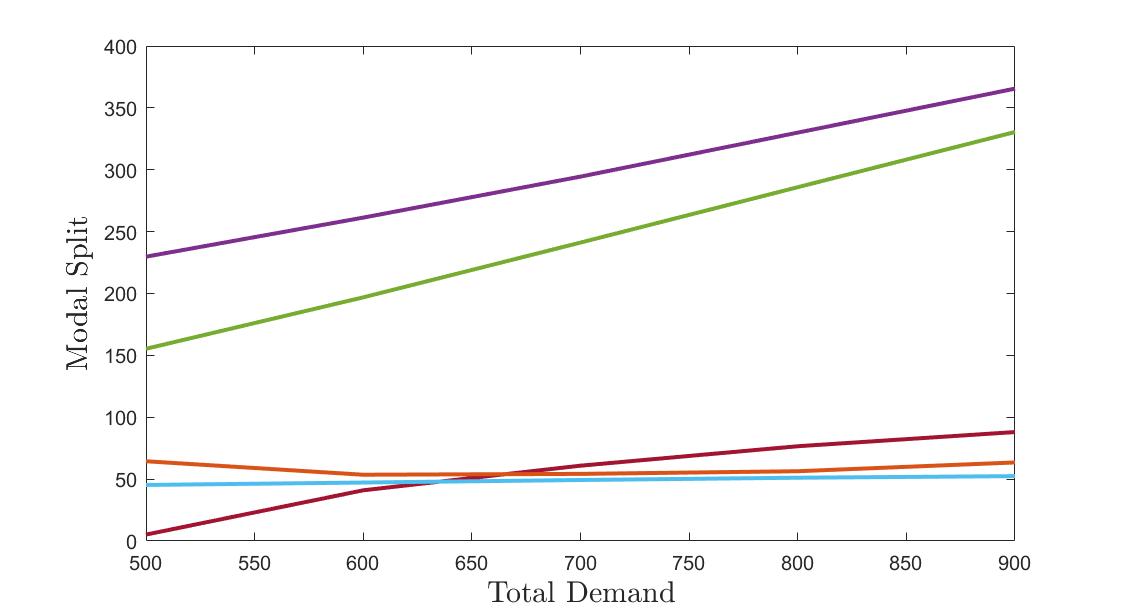}}

  \vspace{0.5cm}
\subfloat[Total Demand: 900]{
  \includegraphics[width=.45\textwidth]{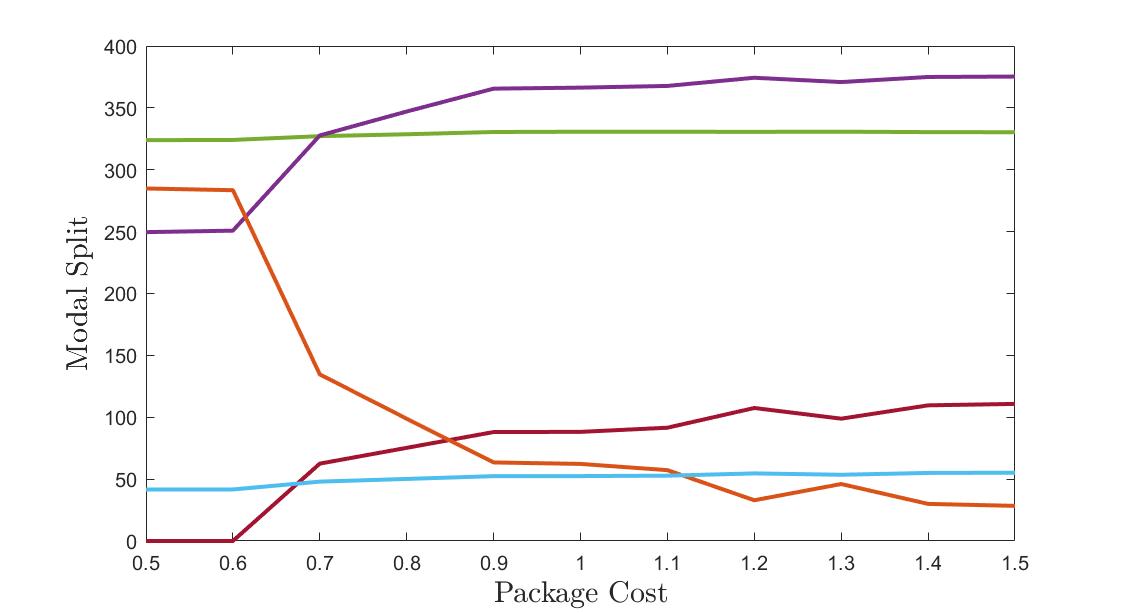}}
  \hspace{0.5cm}
  \subfloat[Package Cost: 1.3]{
  \includegraphics[width=.45\textwidth]{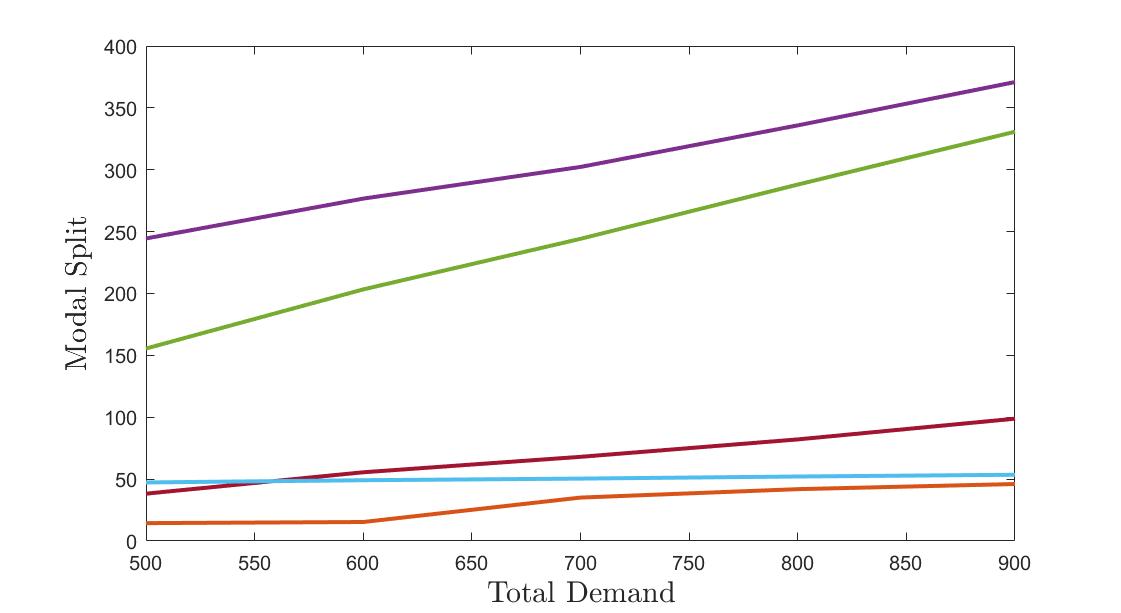}}
  
  \vspace{0.5cm}

  \includegraphics[width=.25\textwidth]{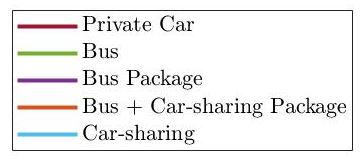}

  \caption{Sections of Figure \ref{Figure 13}}
  \label{Figure21}
\end{figure}

\begin{figure}[h!]
\centering
\includegraphics[scale=0.35]{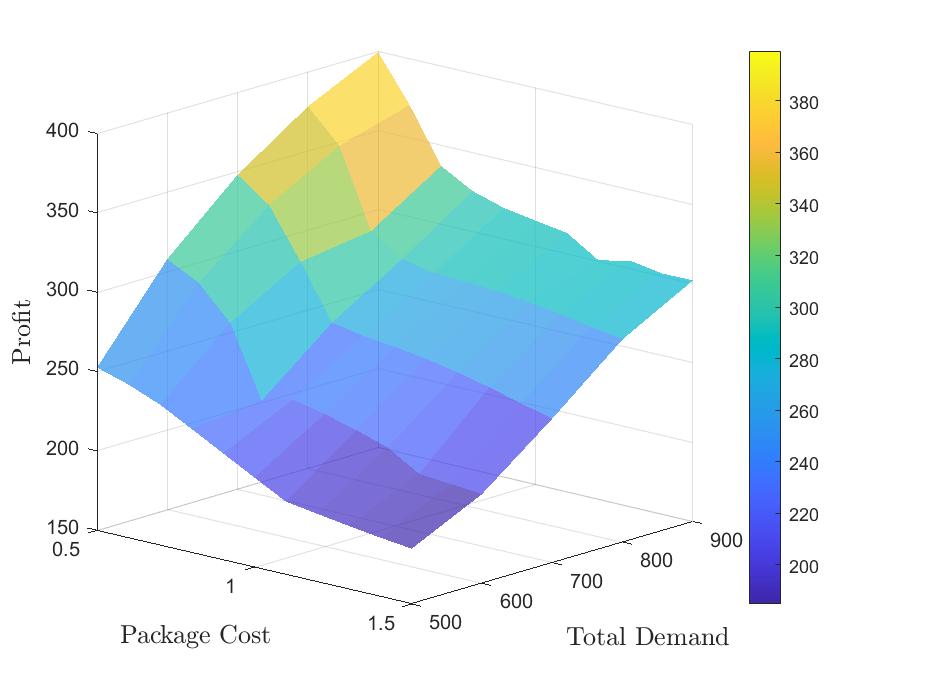}
\caption{Profit variation with package cost} \label{Fig14}
\end{figure}

In Figure \ref{Figure 13} it is clear that, when the Bus + Car-sharing package appears to be more convenient, users tend to subscribe to it, reducing the private car usage. However, as soon as the price increases, we can notice a shift of this behaviour towards the private car. Moreover, the share of users that is subscribing to the car-sharing service remains constant for all the different scenarios. This result could depend on the assumption of fixed fleet size on the car-sharing path. On the contrary, the bus continues to attract most of the demand across the various combinations. In order to have a clearer understanding of the different behaviours with varying price and demand ranges, in Figure \ref{Figure21} we show various sections of Figure \ref{Figure 13}. Specifically, on the left side we show the variation of modal split with the package cost, for fixed values of demand (500,700,900). On the right side, instead, the modal split variation with different values of demand is illustrated based on fixed values of the package cost (0.5 €, 0.9 €, 1.3 €). We can see how for lower values of demand the solutions tend to be more stable, due to the capacity constraints. Bus (green lines) and car-sharing (blue lines) attract the same portion of demand, when the total demand is fixed; users, instead, prefer the to use the combination of bus and car-sharing inside the package only when it is more convenient than bus or private car.

In Figure \ref{Fig14}, instead, we show the variation of profit when the package price changes. This variation follows what was displayed in the results of Figure \ref{Figure 13}: we have the low value of profit when the demand is close to 500 and the package price is at its maximum. Highest values of profit are, instead, registered for the maximum number of subscribers. 

\subsection{Example 3: Competition between MSPs}
\begin{figure}[!tb]
	\centering
	\subfloat[Daily tours]{%
		\includegraphics[width=.9\columnwidth,keepaspectratio]{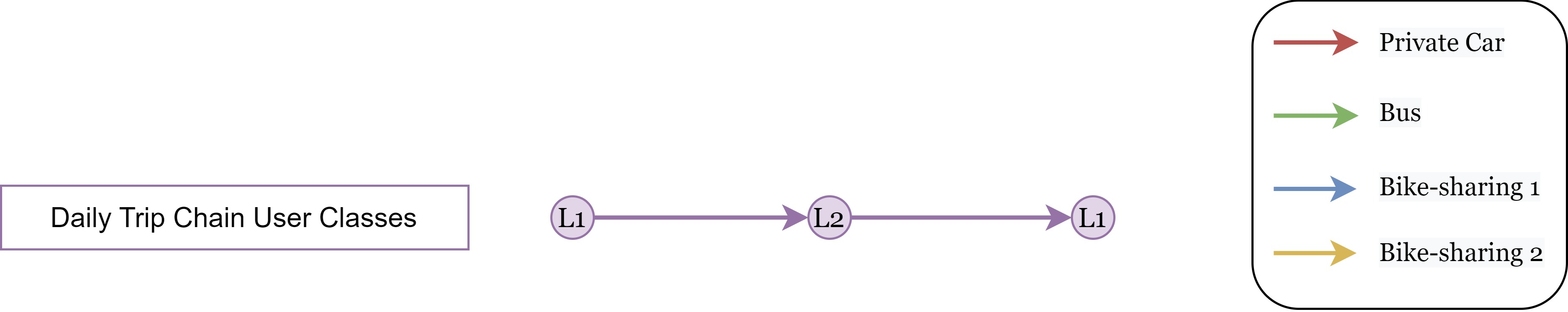}
  \label{Fig20a}}%
 \quad%
	\subfloat[Supernetwork expansion]{%
		\includegraphics[width=.7\columnwidth,keepaspectratio]{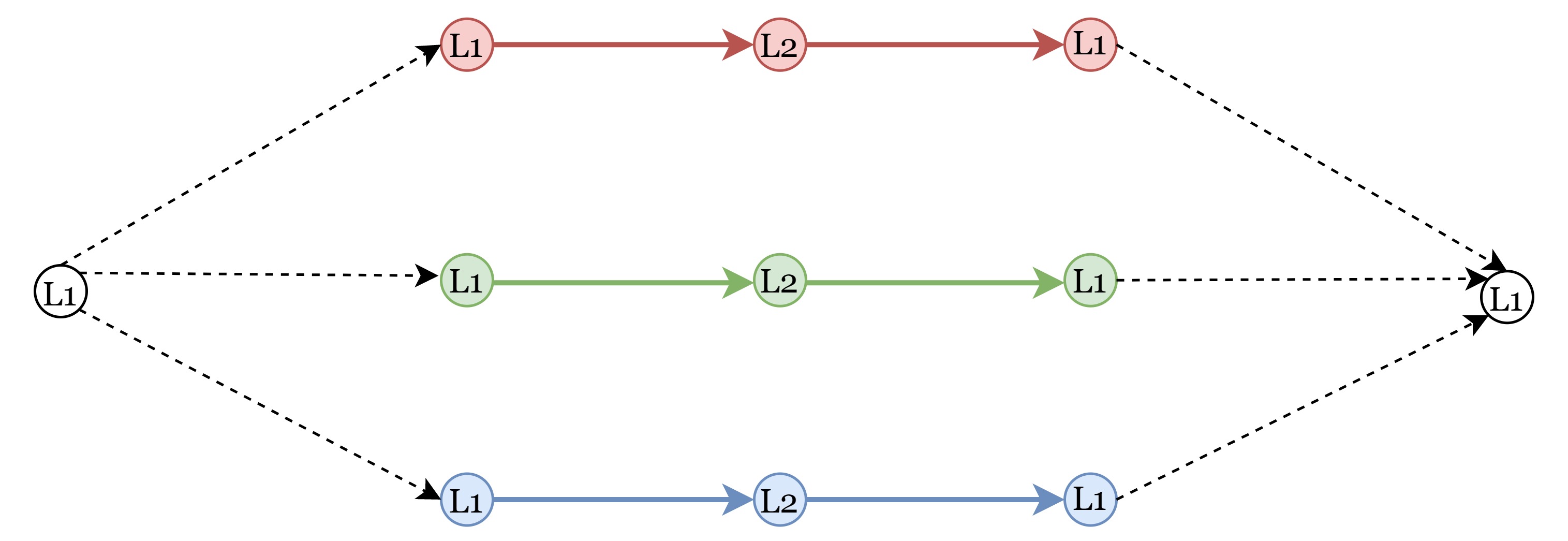}
            \label{Fig20b}%
	}%
 \quad%
	\subfloat[Introduction of a competitor]{%
		\includegraphics[width=.7\columnwidth,keepaspectratio]{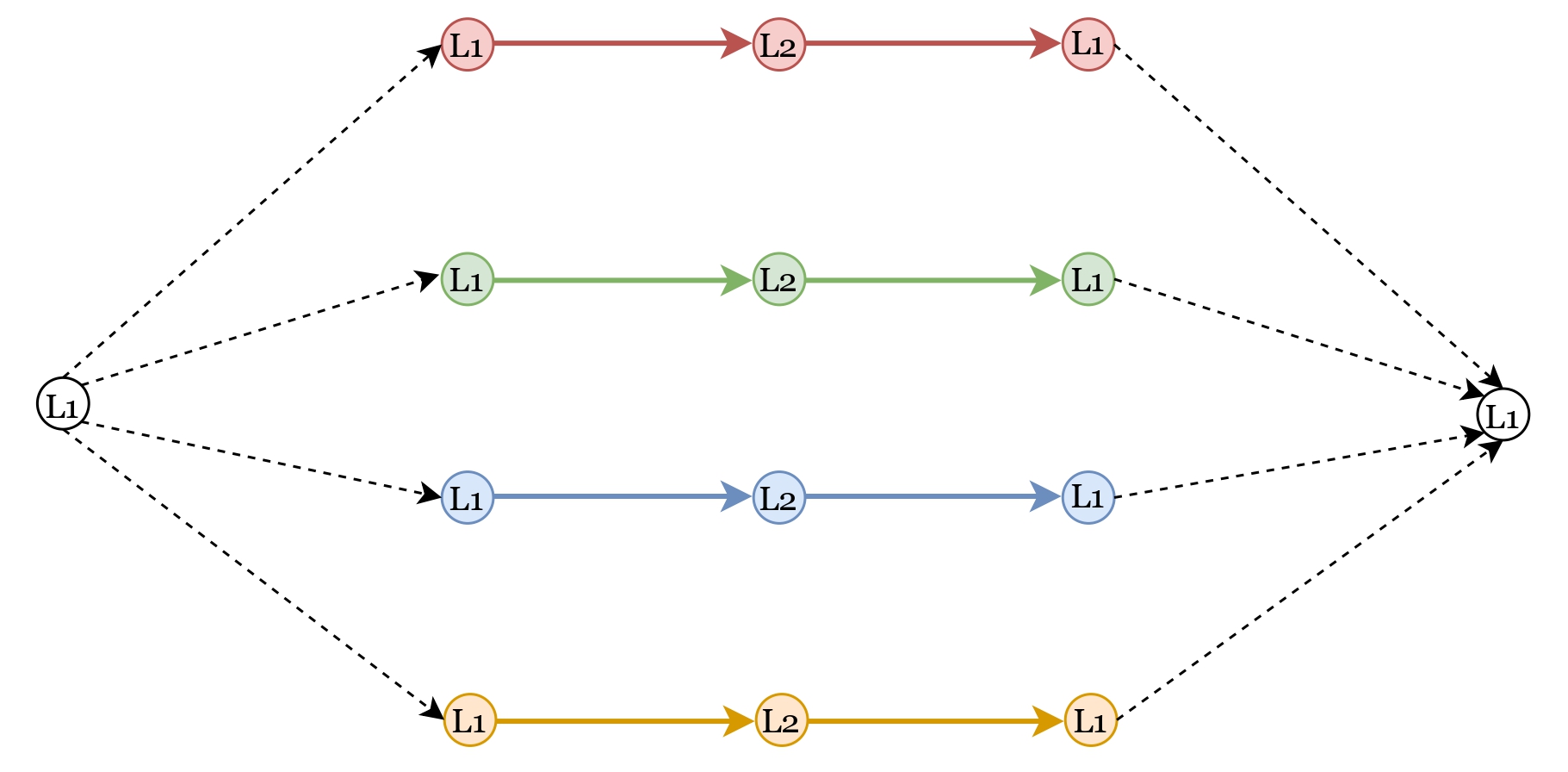}
            \label{Fig20c}%
	}%
    \caption{Example 3}
	\label{Fig20}%
\end{figure}

In the third example we analyse the profitability of a bike-sharing service provider operating in the network displayed in Figure \ref{Fig20}. Due to the reduced distances often covered by bike, we considered a simple trip chain (Figure \ref{Fig20a}) in which two classes of users perform their daily trips. From the network expansion (Figure \ref{Fig20b}), we can see that the available modes of transport are: private car, bus and an electric bike-sharing 1 service. From this configuration, we consider an electric bike-sharing competitor (named bike-sharing 2) entering the market (Figure \ref{Fig20c}) subsidized by a local authority, offering a cheaper service to customers. The service characteristics of the network are defined in Tables \ref{table 5} and \ref{table 6}. 

\begin{table}[ht]
\tbl{Parameters Example}
{
\begin{tabular}{c|c|c|c|c} 
\toprule
 \textbf{Parameters} & \textbf{Private Car} &\textbf{Bus} & \textbf{Bike-sharing 1}& \textbf{Bike-sharing 2}\\ \midrule
$c_s$  & -  &1  &1.5&0.5\\ 
$r_s$  & -  &-  &-&1\\ 
$c^{1}_{a, access}/c^{1}_{a, egress}$  &8 &8.2&10&10 \\
$c^{2}_{a, access}/c^{2}_{a, egress}$  &7.2 & 7.38 & 9&9\\
$c^{1}_{a,wait}$  &- &10 &16&16 \\
$c^{2}_{a,wait}$   &-&9& 14.4&14.4\\
$c^{1}_{a,main}$  & 8&10 &14&14\\
$c^{2}_{a,main}$   &7.2&9&12.6&12.6\\
$c_{a, fuel}$  &0.35&-&-&-\\ 
$c_{a,h}$  &- &- &0.6&1 \\
$l_{a}$  &5 &5 &5&5 \\
$c^{1}_{a,park}$  &9 &- &-&- \\
$c^{2}_{a,park}$  & 8.1&- &-&- \\ 
 \bottomrule
\end{tabular}}
\label{table 5}
\end{table}

\begin{table}[ht]
\tbl{Functional Parameters}
{
\begin{tabular}{c|c|c|c|c|c|c|c|c|c|c|c|c|c|c|c|c} 
\toprule
\textbf{Function} & \multicolumn{5}{c|}{\textbf{Private Car}}  &\multicolumn{5}{c|}{\textbf{Bus}} & \multicolumn{5}{c}{\textbf{Bike-sharing 1/2}}\\ \midrule
 &$t_0$&$\alpha$&$f$& $C$ &$\beta$&$t_0$&$\alpha$&$f$& $C$ &$\beta$&$t_0$&$\alpha$&$f$& $C$ &$\beta$\\\cline{2-16}
$t_{a, access}(f_{a},v_{a})$&\multicolumn{5}{c|}{-} &0.0625&\multicolumn{4}{c|}{-}&0.002& 0.2& $f_a$&$v_a$& 4 \\ \midrule
$t_{a,wait}(f_{a}, v_{a})$&\multicolumn{5}{c|}{-}   &0.25& 0.15&$f_a$ &200&2&0.5& 0.2&$f_a$&$v_a$& 4&\\ \midrule
$t_{a,main}(\textbf{f})$&0.2&4&$\textbf{f}$ &200&2&0.2&4&$\textbf{f}$ &200 &2&0.3&4&$\textbf{f}$ &100& 3&\\\midrule
$t_{a,park}(f_{a}, v_{a})$&0.1& 2.5&$f_a$ & 200&2&\multicolumn{5}{c|}{-} &\multicolumn{5}{c|}{-} \\ \midrule
$t_{a, egress}(f_{a}, v_{a})$ &0.075&\multicolumn{4}{c|}{-} &0.0625&\multicolumn{4}{c|}{-} &0.002& 0.2& $f_a$&$v_a$& 4\\\midrule
$c_{lease}(v)$ &\multicolumn{5}{c|}{-}&\multicolumn{5}{c|}{$0.8v$}&\multicolumn{5}{c}{-} \\
 \bottomrule
\end{tabular}}
\label{table 6}
\end{table}

\begin{figure}[h!]
\centering
\includegraphics[scale=0.3]{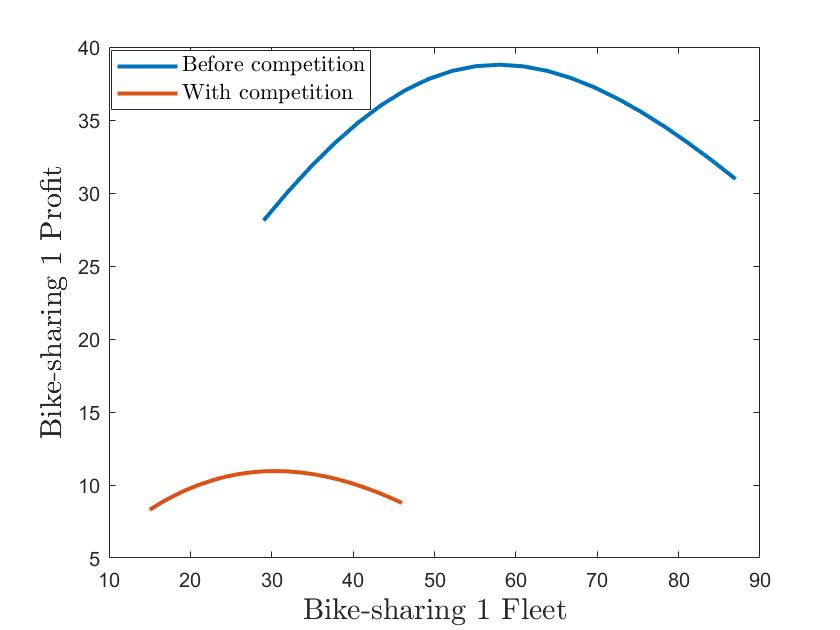}
\caption{Profit variation with fleet size} \label{Fig15}
\end{figure}

\begin{figure}[h!]
\centering
\includegraphics[scale=0.3]{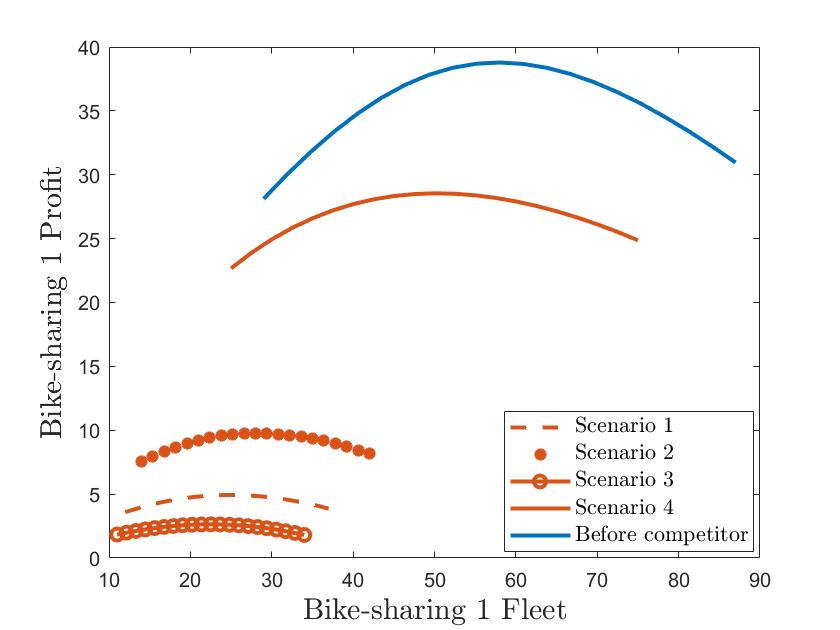}
\caption{Profit variation with fleet size} \label{Fig16}
\end{figure}

The demand of class 1 and 2 respectively is 200 and 120. We considered that the first class has the availability of a private car, instead the service capacity of the bus service is 200, and bike-sharing 2 is 50. In this example, we consider that private car and bus are affecting each other's congestion, and the two bike sharing services are operating in an area in which they share the available bike lanes. 
Applying the proposed methodology before and after the introduction of the competitor, it is possible to see (Figure \ref{Fig15}) a reduction on the maximum value of profit that bike-sharing 1 can obtain when offering the service under the presence of a direct market competitor. For this reason, we define different possible situations that may occur, shown in Figure \ref{Fig16}, to identify applicable strategies to increase the profitability of the service. In Table \ref{table 7} we explicitly described the different situations taken into account.

\begin{table}[h!]
\tbl{Scenarios applicable to Example 3}
{
\begin{tabular}{c|c|c|c} 
\toprule
& \textbf{Package Price}&\textbf{Cost per hour}&\textbf{Subsidies}\\ \midrule
\textbf{Scenario 1}& 1 € & 0.8 & -\\ \midrule
\textbf{Scenario 2}&2 €&-&-\\ \midrule
\textbf{Scenario 3}&0.5 €&1.2&-\\ \midrule
\textbf{Scenario 4}&0.8 €&0.8&1\\
 \bottomrule
\end{tabular}}
\label{table 7}
\end{table}

The tested scenarios show that it is hard for the bike-sharing 1 supplier to reach the same level of profit experienced before the introduction of a competitor. We can see that reducing the price of the package or hourly cost the supplier has to scale down their business. A possible solution to increase the revenues is described by Scenario 4, where we considered that an external supplier subsidize the service, e.g. to sponsor their own business. 

\section{Conclusion}
In order to analyse the relationship between MSPs and users of the transportation network, we defined a MNDP formulated as an MPEC. A general profit maximization formulation has been proposed to describe the behaviour of different MSPs located at the upper level. Users, located at the lower level, are distributed in the network following the traffic network equilibrium conditions. To account for the heterogeneity of user choices in a trip chain-based context, we formulated a multi-class user equilibrium using VI, incorporating non-separable cost functions. We proposed a supernetwork approach to encode various elements such as multimodality, mobility packages, and cooperation/competition between suppliers.

The solution algorithm employed for the MPEC utilized an iterative process to maximize the upper level objective function while applying the MPM to solve the lower level.

In the example section, we presented diverse scenarios to illustrate the potential analysis that could be performed with this methodology. First, we highlighted the significance of non-separability in cost functions to grasp the impact of the co-existence of different transport modes on user choices and estimate the level of service that a supplier can provide. Second, we showcased a mobility package in which a bus provider and car-sharing supplier collaborated to offer a novel service, demonstrating the effects on a supplier's profit based on various pricing strategies. Lastly, we considered the arrival of a new competitor and we proposed different schemes that can be considered to allow the co-existence of all services in the transportation market.

Potential analysis that could be conducted with this methodology include investigating the impact of subsidies by local authorities to encourage public transport usage, introducing additional classes of users with different trip chains, or offering mobility packages with multiple services. 

While we used a small-sized network for our analysis, this methodology has the potential to be scaled up to larger networks in future studies. This proposed methodology represents the first step towards creating a more complex model where multiple suppliers simultaneously optimize their decision variables to assess an equilibrium solution for the entire transportation system.
 
\section*{Disclosure statement}

The authors declare that they have no known competing financial interests or personal relationships with people or organizations that could inappropriately influence their work.

\section*{Funding}

This research is part of the project MaaS4All funded by Fonds National de la Recherche Luxembourg (GRANT NUMBER: 13769009).

\bibliographystyle{tfcad}
\bibliography{biblio}

\end{document}